\documentclass[sn-mathphys,Numbered,pdflatex]{sn-jnl}

\usepackage{graphicx}%
\usepackage{multirow}%
\usepackage{amsmath,amssymb,amsfonts}%
\usepackage{amsthm}%
\usepackage{mathrsfs}%
\usepackage[title]{appendix}%
\usepackage{xcolor}%
\usepackage{textcomp}%
\usepackage{manyfoot}%
\usepackage{booktabs}%
\usepackage{algorithm}%
\usepackage{algorithmicx}%
\usepackage{algpseudocode}%
\usepackage{listings}%

\usepackage{setspace}

\newcommand{\paolo}[1]{\textcolor{black}{#1}}

\newcommand{\typo}[1]{\textcolor{black}{#1}}
\begin{document}

\title[Spatial quantile clustering]{Spatial quantile clustering of climate data}


\author[1]{\fnm{Carlo} \sur{Gaetan}}\email{gaetan@unive.it}
\equalcont{These authors contributed equally to this work.}

\author*[1]{\fnm{Paolo} \sur{Girardi}}\email{paolo.girardi@unive.it}
\equalcont{These authors contributed equally to this work.}

\author[2]{\fnm{Victor Muthama} \sur{Musau}}\email{vmusau@kyu.ac.ke}
\equalcont{These authors contributed equally to this work.}

\affil[1]{\orgdiv{Dipartimento di Scienze Ambientali, Informatica e Statistica}, \orgname{Universit\`a  Ca' Foscari}, \orgaddress{\street{Via Torino 155}, \city{Venezia-Mestre}, \postcode{30172}, \country{Italy}}}

\affil[2]{\orgdiv{Department of Pure and Applied Sciences}, \orgname{Kirinyaga University}, \orgaddress{\street{C74, Kutus- Kerugoya Rd}, \city{Kerugoya}, \postcode{10300}, \country{Kenya}}}

\keywords{Asymmetric Laplace distribution, Markov random field, model-based clustering, time series}


\pacs[MSC Classification]{91C20, 60G60, 62P12, 37M10, 62F15}








\abstract{In the era of climate change, the distribution of climate variables evolves with changes not limited to the mean value. Consequently, clustering algorithms based on central tendency could produce misleading results when used to summarize spatial and/or temporal patterns. We present a novel approach to spatial clustering of time series based on quantiles using a Bayesian framework that incorporates a spatial dependence layer based on a Markov random field. A series of simulations tested the proposal, then applied to the sea surface temperature of the Mediterranean Sea, one of the first seas to be affected by the effects of climate change.}

\maketitle
\section{Introduction}\label{sec1}
\noindent

The extraction of relevant information hidden in complex spatio-temporal climate datasets is one of the main goals of statistical climatology\typo{. C}lustering algorithms are routinely used to summarize and visualize important spatial and/or temporal patterns.

To identify spatial patterns, the most well-known statistical techniques are based on the concept of intra- and inter-cluster variances, such as the k-means algorithm or methods based on empirical orthogonal function analysis.

\typo{K-means is a widely used clustering procedure that identifies clusters by minimizing the distance between each cluster's representative center, called centroid, and the observations it contains.} 
The \typo{distance measure} can be defined using an appropriate norm. This \typo{approach is appropriate} for identifying patterns \typo{related} to unknown mean behaviors. The principle is ideally suited when the variables of interest follow a distribution that spreads around a center, \typo{such} as in the case of a mixture of Gaussian distributions.

On the other hand, \paolo{climate change refers} to long-term shifts in environmental conditions caused primarily by human activities \citep{Schneider:2001}. This phenomenon eventually leads to changes in the distribution of many climatic variables \citep{Portmann:2009, Katz:2010, Sun:2018}. \typo{These changes emphasize the significance of utilizing suitable} statistical methods that take into account no\typo{n-}stationary patterns \typo{when analysing} time series observations \paolo{with a flexible approach. \typo{Quantile regression has gained popularity in detecting changes in various features of climate variable distribution}.  Th\typo{is} method \typo{enables handling of} different quantiles and the presence of outliers \citep{reich:2012, cannon:2018,Vandeskog-et-al:2022}.}

\paolo{Several clustering methods for time series have also been proposed \citep{Liao2005}. 
The former methods are based on raw time series data and adapt distance/similarity measures of multivariate data to time series data.
Latter methods perform a dimension reduction extracting some vectors of features or parameters. The differences between these vectors can then be clustered using standard clustering techniques. Dimension reduction is also \typo{used} in clustering functional data, where the time series is \typo{seen} as a finite realization of a time-varying random function at regular intervals \citep{Zhang:Parnell:2023}.}

Alternatively, we can use model-based clustering techniques \citep{grun2007fitting,Jiang2012}, but the widespread application of the Gaussian distribution may affect the results \paolo{in the presence of asymmetry in the data  distribution.}

\paolo{The presence of outliers in the time series can make it difficult to select the appropriate clustering technique. The robustness of clustering algorithms can be assessed through the application of trimming tools \citep{cuesta1997trimmed,gallegos2002maximum,amovin:2022}. Their main focus was to obtain robust clustering procedures (i.e. robust k-means) developing mathematical methods that perform both classification and outliers detection, as the introduction of restrictions on the dispersion matrices of the groups to avoid detection of spurious clusters \citep{garcia2015avoiding} or the inclusion of constraints on the scatter matrices \citep{fritz2013fast}.}


\typo{Identifying regions with common temporal patterns is one of the potential objectives of clustering climate variables. However, this is complicated by the spatial dimension.
The literature proposes various workarounds to address this issue.}

\paolo{In the \typo{context} of model-based clustering,
 \citet{nguyen2016spatial} modeled the single time series as an autoregressive process, with a Markov random field inducing a spatial structure for the cluster membership.
\citet{disegna2017copula} \typo{employed} a copula-based clustering algorithm. 
Spatial time series can also be treated as spatial functional data \cite{delicado2010statistics,vandewalle2022clustering,koner:2023}.
Still in the framework of model-based clustering \citet{jiang2012clustering} utilized a Markov random field to induce a spatial structure for the cluster memberships of time functions over a \typo{site} network.
This idea has been extended in a Bayesian framework by \citet{hu2022bayesian}.
Instead, \citet{Secchi:Vantini:Vitelli:2013} 
 suggested a bagging-Voronoi strategy for unsupervised classification of functional data.
In a geostatistical context, \citet{Giraldo:Delicado:Mateu:2012} \typo{represented time series as curves using basis functions} and weighted the dissimilarity matrix among the curves using variograms calculated with the coefficients of the basis. 
}

\paolo{
\typo{This paper introduces} a mixture-model approach for time series clustering.
\typo{The focus is on identifying} common long-term trends and seasonal patterns across a collection of time series of sea surface temperature measurements on a grid that spans the Mediterranean Sea.}
The dataset is particularly challenging due to the skewness of the data distribution, time-varying heteroscedasticity, and the presence of outliers.
\typo{Our model considers time-varying quantiles, enabling us to address the aforementioned issues and potentially attain more reliable clustering outcomes.} 

\paolo{To model a single time series at a grid point, we exploit the connection between asymmetric Laplace regression and quantile regression \cite{bera:2016}.
\typo{Our regression model is additive and includes a smooth component that represents the multi-year trend in the time series, as well as modulated harmonic components that estimate seasonal variation.}
\typo{We incorporated} spatial dependence over the grid points into the model through a Potts model \citep{Potts:1952}. The Bayesian inference for the resulting hierarchical model is based on Markov Chain Monte Carlo (MCMC) methods. }

The article is \typo{structured} as follows. In the next section (Section 2), we provide \typo{additional} details about the data.  Section 3 introduces the proposed model-based clustering procedure, and Section 4 presents a simulation study to illustrate the performance and peculiarities of the \typo{proposed} procedure. Section 5 collects the results of the application to the Mediterranean Sea surface temperature data. In Section 6, we \typo{analyze} the relative strengths and weaknesses of our proposal.
\section{Real data}\label{sec2}

Surrounded by Europe, Africa, and Asia, the Mediterranean Sea is the largest semi-enclosed European sea. The strait of Sicily separates the Western Mediterranean and the Eastern Mediterranean, the major basins, while other sub-basins such as the Alboran, Ligurian, Tyrrhenian, Adriatic, Ionian, and Aegean Seas fragment the area in specific zones of interest. The Mediterranean Sea is \typo{connected} to other three \typo{bodies of water}: the Atlantic Ocean, the Black Sea, and the Red Sea through the Strait of Gibraltar, the Dardanelles, and the Suez Canal, respectively. The presence of different conditions passing from shallow waters to deep ones permits us to consider the Mediterranean Sea as a scale model of the world’s oceans in terms of water circulation and deep water formation \citep{nykjaer:2009,lejeusne:2010}. 

\begin{figure}
	\centering
	\begin{tabular}{c}
\includegraphics[width=1\linewidth]{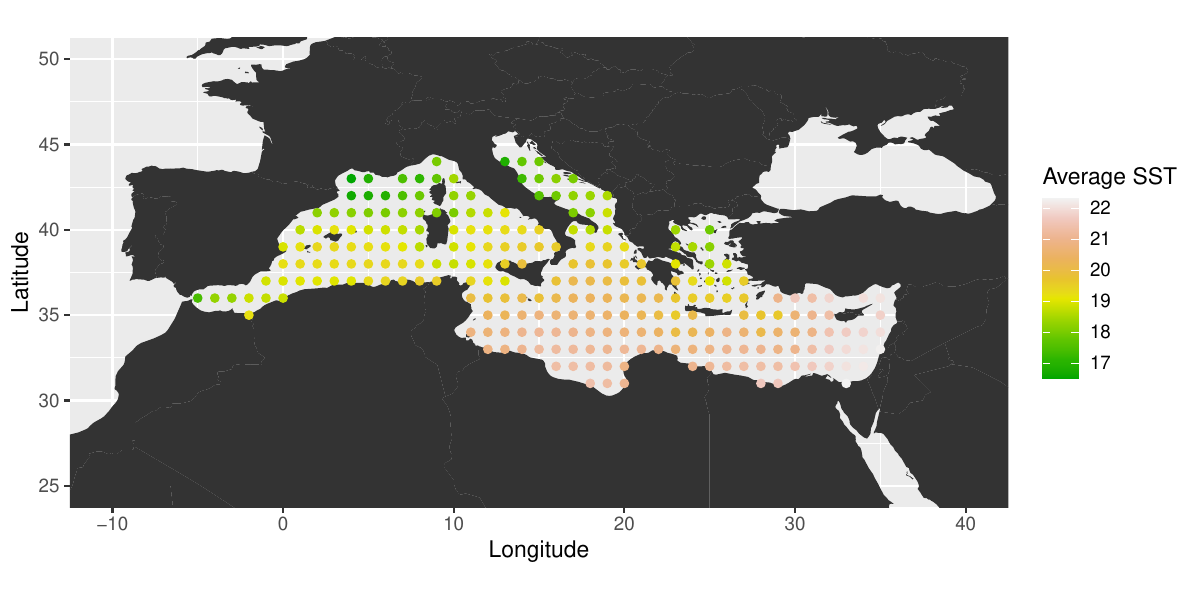} 
	\end{tabular}
	\caption{Average \paolo{Sea Surface Temperature (SST)} of the Mediterranean Sea over the period 1982 - 2012.}\label{fig:1}
\end{figure}

The Mediterranean Sea was one of the first world seas where temperature changes \typo{could} be attributed to climate change \citep{bethoux:1990}. Air temperature is strongly influenced by atmospheric phenomena, not even connected to global warming, but the increase of heat waves in the Mediterranean Sea \citep{ibrahim:2021} affected the marine plant and animal world, in particular, the maintenance of biodiversity \citep{nunes:2019}. However, the Sea Surface Temperature (SST) is a \typo{significant} indicator of climate change, since it can, in turn, \typo{impact atmospheric circulation}. Normally the SST is influenced by various factors, \typo{including}  solar irradiation \typo{intensity},   water bathymetry,  presence of superficial and deep currents, and the presence of rivers.
The recent history of SST in the Mediterranean Sea shows a slow decrease in the upper layer (150 m) until the 1980s and then a significant increase of about 0.4$\,^o$C  \typo{until} 2015 \citep{pastor:2019}.

We have considered a dataset from a reanalysis of SST consisting of monthly mean SST in Celsius degrees from January 1982 to December 2012 for a total of $372$ time points
 (see \url{https://data.marine.copernicus.eu/product/SST_MED_SST_L4_REP_OBSERVATIONS_010_021/description} for a complete data description). Gridded data is available at 1-degree resolution for each month, with a total of $251$ grid points extending to the entire Mediterranean Sea. 
 There are no missing values in the data set.
\begin{figure}
	\centering
	\begin{tabular}{cc}
	\includegraphics[width=0.95\linewidth]{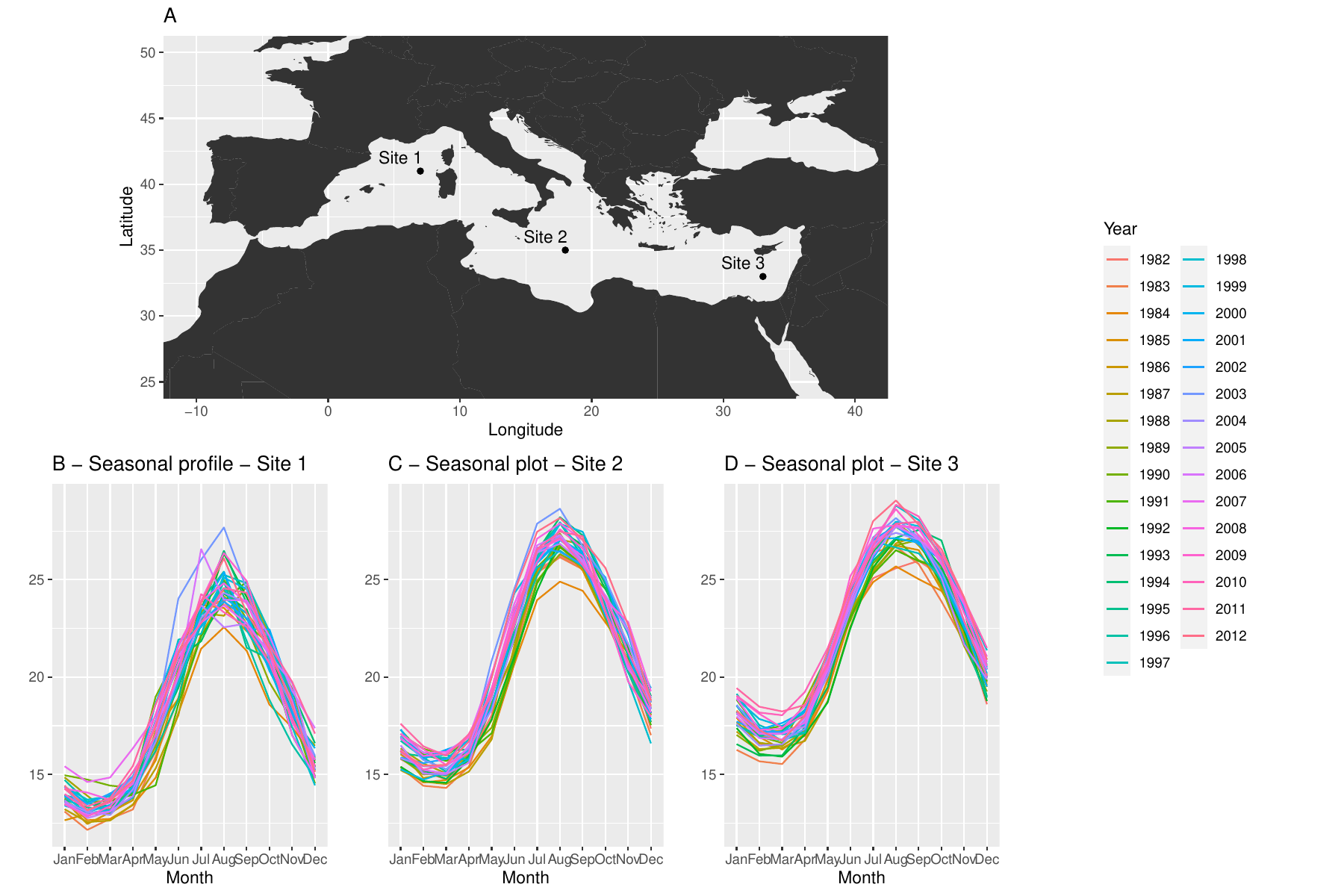}
	\end{tabular}
	\caption{Three sites selected on the Mediterranean Sea (A) and seasonal profile (B, C, and D).}\label{fig:2}
\end{figure}

In Figure \ref{fig:1} we \typo{show} the overall mean SST for all sites. The mean SSTs range between 16.5$\,^o$C and 22.3$\,^o$C with a northwest-southeast gradient and, in particular, the coldest water masses are recorded near the Gulf of L\paolo{i}on, while the warmest ones belong to the southern part of the Levantine Sea, near the Egyptian coastal area.
As reported by Figure \ref{fig:2}, among three selected sites (selected following an NW to SE direction) a strong seasonal pattern is evident as well as the presence of different temperature extremes. In addition, the highest values were recorded \typo{at all the three sites during the years 2006 to 2012, especially at} site 3 located in the Levantine Sea between Egypt and Cyprus, indicating \typo{a general increase in SST}.

\begin{figure}
	\centering
	\includegraphics[width=0.8\linewidth]{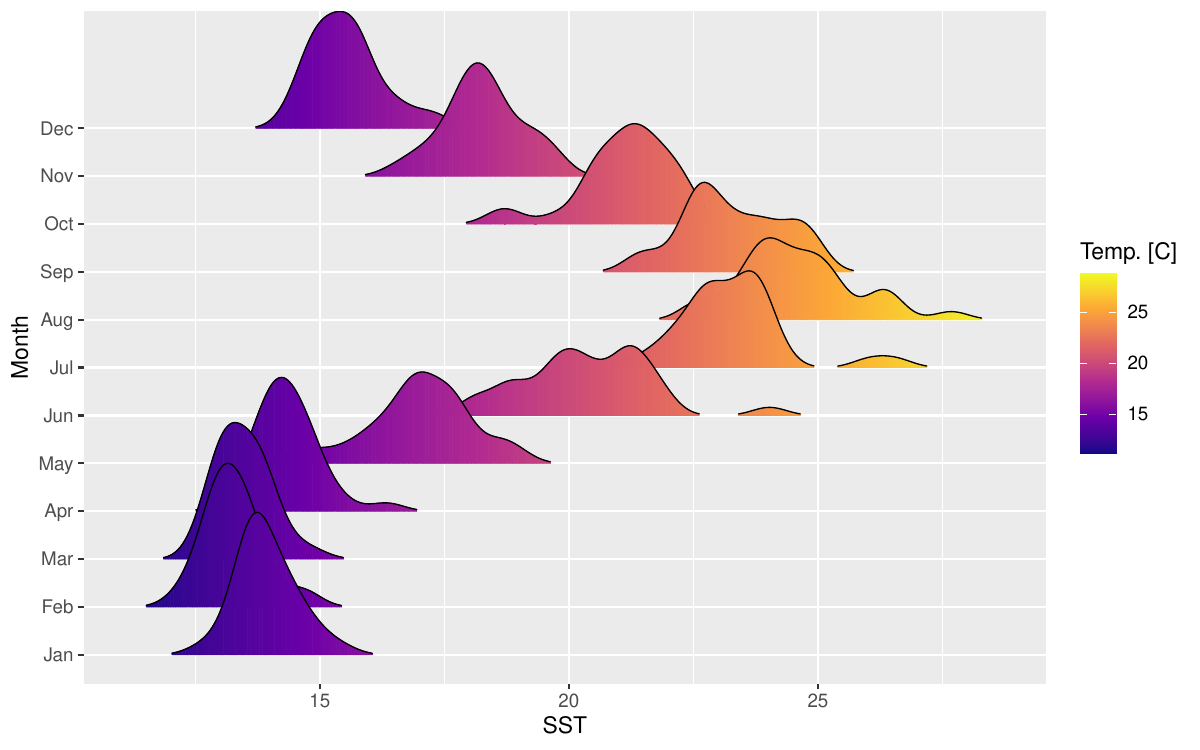}
	\caption{Average SST distribution by month of observation in the Algerian Basin.}\label{fig:3}
\end{figure}

The clustering of the different Mediterranean areas seems difficult not only because of the presence of a shaded increasing trend and a strong seasonality but also because of the asymmetric distribution of the SST, as reported for site 3 in the Algerian Basin (Figure \ref{fig:3} - D).

\typo{Classifying} SST behavior using robust approaches and considering a specific quantile of interest (e.g., 90th percentile) can help identify areas with different trends and seasonal patterns. 
\section{Mixture-model clustering}\label{sec3}

We consider a finite set of sites $\mathcal{S} = \{1, \ldots, N\}$ . Each site is associated with a given spatial localization. At each site $i \in \mathcal{S}$, we observe a time series of $T$ observations, $\boldsymbol{y_i}=(y_{i,1},\ldots,y_{i,T})'$. The goal is to cluster the $N$ time series into $K$ groups to detect homogenous regions.

We follow a  mixture-model approach \citep{mclachlan:peel:2000}  to clustering, according to which the cluster membership of the $i$-th time series is represented by a latent random variable  ${c}_i \in \{1,\ldots,K\} $, where $c_{i} = k$ indicates that the $i$-th time series belongs to the group $k$. 
\subsection{Membership model}
To model the memberships,  we suppose that $\boldsymbol{c}=(c_1,\ldots,c_N)^\prime$ are independent and identically distributed multinomial variables  with probability mass function for 
\paolo{
\begin{eqnarray}\label{eq:ind}
	f(\boldsymbol{c}|\boldsymbol{\alpha})&=&
\prod_{i=1}^N\frac{\operatorname{exp}\left(\sum_{k=1}^{K}\alpha_{k}I(c_{i}=k)\right)}{\exp\left(\sum_{k=1}^{K}\alpha_{k}\right)}=
 \frac{\operatorname{exp}\left(\sum_{k=1}^{K} \alpha_{k}n_k\right)}{\exp\left(N\sum_{k=1}^{K}\alpha_{k}\right)}\nonumber\\
&=&
 \frac{\operatorname{exp}\left(\sum_{k=1}^{K} \alpha_{k} n_k\right)}{Z(\boldsymbol{\alpha})}
\end{eqnarray}
where $n_k=\sum_{i=1}^N I(c_{i}=k)$,  
$Z(\boldsymbol{\alpha})= \exp\left(N\sum_{k=1}^{K}\alpha_{k}\right)$, with   $\boldsymbol{\alpha}=(\alpha_1,\alpha_2,\ldots,\alpha_K)^\prime$, and  $\alpha_1=0$ for model identifiability, since $N=\sum_{k=1}^Kn_k$.}

However, when time series are collected over a spatial domain, cluster memberships are characterized by some dependence. In this paper, the spatial dependence among site memberships on a network is modeled using a Markov Random Field (MRF).

We denote \typo{with} $i\sim l$ if  $i$ and $l$ are neighbors
and the set $\partial i=\{l: i\sim l\}$ indicates the neighborhood of $s_i$.  A MRF is specified in terms of a neighborhood system, i.e., the set of the neighborhood $\mathcal{G}=\{\partial i,\quad i=1,\ldots,n\}$.
  The Potts model \citep{Potts:1952,Strauss:1977} is a MRF with probability mass function
$$ 	f(\boldsymbol{c}|\boldsymbol{\alpha},\boldsymbol{\beta})= \frac{\exp\left(\sum_{k=1}^K  \alpha_k n_k +
\sum_{i\sim j} \sum_{k\le l} \beta_{kl} I(c_i=k)I(c_j=l)\right)}{Z(\boldsymbol{\alpha},\boldsymbol{\beta})}
$$
with $\boldsymbol{\beta}=(\beta_{11},\beta_{12},\ldots,\beta_{1K},\beta_{22},\beta_{23},\ldots,\beta_{2K},\ldots,\beta_{(K-1)K})^\prime$.
Here $\sum_{i\sim j}$ means that the summation is taken over all the pairs of neighboring sites.
The parameter $\boldsymbol{\beta}$ controls the spatial dependence, allowing a spatial dependence inside each cluster and spatial interaction between two clusters.

Note that, for a moderate or large number of sites, the calculation of the normalizing constant  
$$
Z(\boldsymbol{\alpha},\boldsymbol{\beta})=\sum_{c_1=1}^K\cdots \sum_{c_n=1}^K \exp\left(\sum_{k=1}^K   \alpha_k n_k +
\sum_{i\sim j} \sum_{k\leq l} \beta_{kl} I(c_i=k)I(c_j=l)\right)
$$
is intractable because it involves $K^N$ terms.

\paolo{In this paper, we examine a simplified variation of the Potts model \citep{jiang2012clustering, Gaetan:2017}. We achieve this by setting $\beta_{kl}=0$ for $k\ne l$ and $\beta_{kk}=\beta_k$. 
\begin{eqnarray}\label{eq:Potts}
f(\boldsymbol{c}|\boldsymbol{\alpha},\boldsymbol{\beta})
&=& \frac{\exp\left\{\sum_{k=1}^K \left(\alpha_k n_k +
\beta_k \sum_{i\sim j}   I(c_i=k)I(c_j=k)\right)\right\}}{Z(\boldsymbol{\alpha},\boldsymbol{\beta})}\nonumber
\\
&=& \frac{\exp\left\{\sum_{k=1}^K  \left(\alpha_k n_k+
	\beta_k\sum_{i=1}^N n_{k,i}/2\right)\right\}}{Z(\boldsymbol{\alpha},\boldsymbol{\beta})},\end{eqnarray}
where $n_{k,i}=\sum_{j \in \partial_i} I(c_i=k)I(c_j=k)$ is the number of the neighbors of the site $i$ with  membership equal to $k$.
Consequently, we explore distinct spatial dependence parameters ($\beta_k$) in each cluster. The parameter $\beta_k$ can be interpreted as a regularization parameter where higher values of $\beta_k$ favor clustering to cluster $k$. By setting $\beta_k=0$ for all $k$}, model \eqref{eq:Potts} degenerates to the independent model \eqref{eq:ind}. 

\subsection{Time series model}

We are mainly interested in clustering the time series according to their spatio-temporal pattern, so we model the $i$-th time series 
by 
$y_{i,t} = \mu_{c_i}(t) + e_{i,t},$ 
where the deterministic function $\mu_{c_i}(t)$  represents trend and/or seasonality for each group, and $e_{i,t}$ represents an additive independent error.
A flexible specification of  $\mu_{c_i}(t)$ is given  by a linear combination of \paolo{known} functions, so that
\begin{equation}\label{eq:regression}
\mu_{c_i}(t)=\sum_{j=0}^J\gamma_{c_i,j} \psi_j(t)=\boldsymbol{\psi}(t)^\prime\boldsymbol{\gamma}_{c_i}    
\end{equation}
where $\boldsymbol{\psi}(t)=(\psi_0(t),\ldots,\psi_J(t)^\prime$ and $\boldsymbol{\gamma}_{c_i}=(\gamma_{c_i,0},\ldots,\gamma_{c_i,J})^\prime$.  In this way, we result in a different \paolo{parametric} linear model for each group. 
The parameters $\gamma_{c_i,0}$ can introduce an offset in the data, by setting the 
function $\psi_0(t) = 1$. There are many possible choices
for the functions: polynomial,  kernel, sigmoidal, spline, trigonometric, wavelet, etc. 
A common assumption in clustering time series is that the additive error $e_{i,t}$ is symmetric and  Gaussian distributed with mean $0$ and standard deviation $\sigma_{c_i}$.
This implies that $\mu_{c_i}(t)=\boldsymbol{\psi}(t)^\prime\boldsymbol{\gamma}_{c_i}$ is the mean regression function, which is fitted by evaluating a quadratic loss function, i.e. $(y_{i,t}-\boldsymbol{\psi}(t)^\prime\boldsymbol{\gamma}_{c_i})^2$, in the likelihood. However, we could be interested in considering  a loss function
 with more robust properties or modelling different parameters of the marginal distribution of the time series.
 
 Here we follow the regression quantile approach \citep{koenker1978regression} and assume that $e_{i,t}$ has the density function $d (\cdot| p)$ and $p$-th, $0<p<1$, quantile equal to zero, that is, 
 $\int_{-\infty}^0 d (e| p)de = p$. It turns out that the $p$-th  quantile of $y_{i,t}$ is   simply
 $
 Q(y_{i,t})=\boldsymbol{\psi}(t)^\prime\boldsymbol{\gamma}_{c_i}
 $.
 In particular, we assume that  $e_{i,t}$ has Asymmetric Laplace Distribution,   $\operatorname{ALD}(p,\sigma)$, \citep{koenker1999goodness} with density
 \begin{equation}
	f(e|\sigma^2)=\frac{p(1-p)}{\sigma}\exp\left\{-\rho_{p}\left(\frac{e}{\sigma}\right)\right\}\label{eq:ald}
\end{equation}
where $ \rho_{p}(\cdot) $ is the  loss function $ \rho_{p}(u)=u(p-I(u<0)) 
$  and  $ \sigma>0 $.
Therefore, the density of $\boldsymbol{y}_i$ given $c_i$ is
 \begin{eqnarray}
	f(\boldsymbol{y}_i|c_i,\boldsymbol{\gamma}_{c_i},\sigma_{c_i}^2)&=&\prod_{t=1}^T f({y}_{it}|\boldsymbol{\gamma}_{c_i},\sigma_{c_i}^2)\nonumber\\
	&=&\prod_{t=1}^T
	\frac{p(1-p)}{\sigma_{c_i}}\exp\left\{-\rho_{p}\left(\frac{y_{i,t}-\boldsymbol{\psi}(t)^\prime\boldsymbol{\gamma}_{c_i}}{\sigma_{c_i}}\right)\right\}\nonumber\\
	&=&\left\{\frac{p(1-p)}{\sigma_{c_i}}\right\}^T\exp\left\{-\sum_{t=1}^T\rho_{p}\left(\frac{y_{i,t}-\boldsymbol{\psi}(t)^\prime\boldsymbol{\gamma}_{c_i}}{\sigma_{c_i}}\right)\right\}\label{eq:fy}
\end{eqnarray}
with $\boldsymbol{y}_i=(y_{i1},\ldots,y_{iT})^\prime$.
 For example, when $p=1/2$, the conditional median regression is more robust than the conditional mean regression with outliers in the observations \citep{Huber:1981}.
\section{Model estimation and inference}
We  consider the full set of observations $\boldsymbol{y}=\{\boldsymbol{y}_{1}^\prime,\ldots,\boldsymbol{y}_{N}^\prime\} $ 
and we want to make inference on the model parameters $\boldsymbol{\theta}=(\boldsymbol{\alpha},\boldsymbol{\beta},\boldsymbol{\gamma}^\prime,\boldsymbol{\sigma^2}^\prime)'$, with $\boldsymbol{\gamma}=(\boldsymbol{\gamma}_1^\prime,\ldots,\boldsymbol{\gamma}_K^\prime)^\prime$ and $\boldsymbol{\sigma^2}=(\sigma^2_1,\ldots,\sigma^2_K)$.
The joint distribution of $\boldsymbol{y},\boldsymbol{c}$ is given by
$$
f(\boldsymbol{y},\boldsymbol{c}|\boldsymbol{\theta})=\prod_{i=1}^N f(\boldsymbol{y}_i|c_i,\boldsymbol{\gamma}_{c_i},\sigma^2_{c_i})f(\boldsymbol{c}|\boldsymbol{\alpha},\boldsymbol{\beta}),
$$
and the likelihood is 
$$
f(\boldsymbol{y}|\boldsymbol{\theta})=\sum_{\boldsymbol{c}}\prod_{i=1}^Nf(\boldsymbol{y}_i|c_i,\boldsymbol{\gamma}_{c_i},\sigma^2_{c_i})f(\boldsymbol{c}|\boldsymbol{\alpha},\boldsymbol{\beta})
$$
Its evaluation is computationally demanding due to the normalizing constant $Z(\boldsymbol{\alpha},\boldsymbol{\beta})$.
A workaround
 relies on a pseudo-likelihood approximation \citep{Besag:1986} of $f(\boldsymbol{c}|\boldsymbol{\alpha},\boldsymbol{\beta})$, namely
 $$f^*(\boldsymbol{c}|\boldsymbol{\alpha},\boldsymbol{\beta})=\prod_{i=1}^Nf(c_i|c_j,j\sim i,\boldsymbol{\alpha},\boldsymbol{\beta}),
 $$
 where the conditional distribution for a location $i$, due to the Markov property of Potts model, is given by
 \paolo{\begin{equation}\label{eq:condPotts}
	f(c_i|c_j, j\sim i ,\boldsymbol{\alpha},\boldsymbol{\beta})=	
	\cfrac{\exp(\alpha_{c_i} +
		\beta_{c_i} n_{c_i,i})}
	{\sum_{k=1}^K 
		\exp(\alpha_{k} + 	\beta_k n_{k,i})}.
\end{equation}
}

This approach has been criticized in the literature \citep{cucala_et_al:2009} despite the reduction in complexity.
In the following, we follow a Bayesian approach to estimation and first specify a prior distribution $\pi(\boldsymbol{\theta})$ for the parameter $\boldsymbol{\theta}$.
\subsection{Prior distribution}
We assume  independent prior distributions for each component of $\boldsymbol{\theta}$, namely
\begin{equation}\label{eq:prior}
	\pi(\boldsymbol{\theta})=\pi(\boldsymbol{\alpha})\times \pi(\boldsymbol{\beta})\times \prod_{k=1}^K\pi(\boldsymbol{\gamma}_{k})\times
	\prod_{k=1}^K\pi(\sigma^2_{k}).
\end{equation}
and we remember that we have imposed $\alpha_1=0$ in the equation
\eqref{eq:Potts}.
 The  prior distribution for the remaining parameters $\boldsymbol{\alpha}$ is the $K-1$ multivariate Gaussian distribution $\mathcal{N}(\boldsymbol{0},a \boldsymbol{I})$ with an improper prior for   $a\gg 0$.
The prior for each $\beta_k$ is a uniform distribution on $(0, B)$, where $B$ is the maximum possible value of $\beta_k$.
\paolo{Note that our a priori preference is for the spatial aggregation.
The value of $B$ can greatly influence the spatial aggregation.
It should be noted that when the length of the time series increases, the likelihood term in the a posteriori distribution formula for the memberships becomes predominant defeating the role of aggregation. As a consequence, if the model is used to favor spatial aggregation and pattern simplification, the value of $B$ should be adjusted according to the length of the time series. 
In fact, in numerical examples where the time series have different lengths, we use two values for $B$. Specifically, $B=10$ for the simulation experiments and $B=100$ for the real data case.}

We assume that \paolo{the coefficients  $\boldsymbol{\gamma}_{k}$,  $k=1,\ldots,K$  in \eqref{eq:regression}} are independent and identically distributed according to the $K$ multivariate Gaussian distribution $\mathcal{N}(\boldsymbol{0},g \boldsymbol{I})$, where $g\gg0$.
Finally, similar to \citet{Benoit2017}, we propose the inverse gamma distribution $\mathrm{IG}(s_{0},d_{0})$ as the prior for each $\sigma_k$, where the shape and scale parameters $s_{0}$ and $d_{0}$, respectively, are known.

\subsection{Hybrid Gibbs sampler}

 The evaluation of the posterior distribution $\pi(\boldsymbol{\theta}|\boldsymbol{y})\propto 
f(\boldsymbol{y}|\boldsymbol{\theta})\pi(\boldsymbol{\theta})$ requires an implementation of a  Markov Chain Monte Carlo (MCMC) algorithm \citep{Robert2004}.
The popular Gibbs sampler successively draws samples according to the full conditional distributions associated with the distribution of $\pi(\boldsymbol{\theta}|\boldsymbol{y})$. When a conditional distribution cannot be easily sampled, one can resort to a Metropolis-Hastings (MH) move, which generates samples according to an appropriate proposal and accept or reject these generated samples with a given probability. The MCMC algorithm is called hybrid Metropolis-within-Gibbs  sampler  and generates samples that are asymptotically distributed according to the posterior distribution.

To develop our MCMC algorithm  we  exploit  a mixture representation
\citep{Kotz2001,kozumi2011gibbs} of $e_{i,t}$ in the equation \eqref{eq:ald} for sampling from the  conditional distributions.
This trick is known as completion \citep[][Definition 10.3]{Robert2004}. More specifically, let $\nu_{it}$ and $w_{it}$ be mutually independent random variables, with $\nu_{it}$ being a standard Gaussian random variable and $w_{it}$ being a standard exponential random variable. 
Given $c_i$, $y_{it}$ can be rewritten as
\begin{eqnarray}
	y_{it}=\boldsymbol{\psi}(t)^\prime\boldsymbol{\gamma}_{c_i}+\tau\sigma_{c_i} w_{it}+\omega\sigma_{c_i} \sqrt{w_{it}}\,\nu_{it} \label{eq3}
\end{eqnarray}
where $\tau=(1-2p)/\{p(1-p)\}$, $\omega^2={2}/\{p(1-p)\}$.

The joint density of $ \boldsymbol{y}_i$ given $ \boldsymbol{w}_i=(w_{i1},...,w_{iT})' $ and $c_i$ is
\begin{eqnarray}
	f(\boldsymbol{y}_i|c_i,\boldsymbol{\gamma}_{c_i},\sigma^2_{c_i},\boldsymbol{w}_i)\propto {(\sigma^2_{c_i})}^{-{T}/{2}}\prod_{t=1}^{T} w_{it}^{-{1}/{2}}\exp\left\{-\frac{1}{2}\sum_{t=1}^{T}\frac{(y_{it}-\boldsymbol{\psi}(t)^\prime\boldsymbol{\gamma}_{c_i}-\tau\sigma_{c_i} w_{it})^{2}}{\sigma^2_{c_i}\omega^{2}w_{it}}\right\} \nonumber\\
 \label{eq5}
\end{eqnarray}

Therefore instead of devising a MCMC algorithm
for sampling from $\pi(\boldsymbol{\alpha},\boldsymbol{\beta},\boldsymbol{c},\boldsymbol{\gamma},\boldsymbol{\sigma^2}|\boldsymbol{y})$, we devise an algorithm for sampling from
  \begin{eqnarray*}
 	\pi(\boldsymbol{\alpha},\boldsymbol{\beta},\boldsymbol{c},\boldsymbol{\gamma},\boldsymbol{\sigma^2}, \boldsymbol{w}|\boldsymbol{y})&\propto&\prod_{i=1}^{N} 
 	f(\boldsymbol{y}_i|c_i,\boldsymbol{\gamma}_{c_i},\sigma^2_{c_i}, \boldsymbol{w}_i)f(\boldsymbol{w}_{i})
 	\times f(\boldsymbol{c}|\boldsymbol{\alpha},\boldsymbol{\beta})\times \nonumber\\
 	&&\prod_{k=1}^K\pi(\boldsymbol{\gamma}_{k})\times \pi(\sigma^2_{k}) \times\pi(\boldsymbol{\alpha})\times \pi(\boldsymbol{\beta}) 
 \end{eqnarray*}
 with $\boldsymbol{w}=(\boldsymbol{w}_1^\prime,\ldots, \boldsymbol{w}_N^\prime)^\prime$. 
 Our MCMC algorithm  is described below (Algorithm 1).
 
 \bigskip

\begin{algorithm}
\caption{Hybrid MCMC algorithm}\label{eq:alg1}
\begin{algorithmic}[1]
\Require initial $\boldsymbol{\alpha}^{(0)},\boldsymbol{\beta}^{(0)},\boldsymbol{c}^{(0)},\boldsymbol{\gamma}^{(0)},\boldsymbol{\sigma^2}^{(0)},\boldsymbol{w}^{(0)}$, number of iterations $M$
\State $m \leftarrow 1$
\While{$ m  \leq M$ }    
\State Generate $\boldsymbol{\gamma}^{(m)}$ from $\pi(\boldsymbol{\gamma}|\boldsymbol{\alpha}^{(m-1)},\boldsymbol{\beta}^{(m-1)},\boldsymbol{c}^{(m-1)},\boldsymbol{\sigma^2}^{(m-1)},\boldsymbol{w}^{(m-1)},\boldsymbol{y})$\
\State Generate $\boldsymbol{\sigma^2}^{(m)}$ from $\pi(\boldsymbol{\sigma^2}|
	\boldsymbol{\alpha}^{(m-1)},\boldsymbol{\beta}^{(m-1)},\boldsymbol{c}^{(m-1)},\boldsymbol{\gamma}^{(m)},\boldsymbol{w}^{(m-1)},\boldsymbol{y})$\
\State Generate $\boldsymbol{w}^{(m)}$ from $\pi(\boldsymbol{w}|
	\boldsymbol{\alpha}^{(m-1)},\boldsymbol{\beta}^{(m-1)},\boldsymbol{c}^{(m-1)},\boldsymbol{\gamma}^{(m)},\boldsymbol{\sigma^2}^{(m)},\boldsymbol{y})$\
\State Generate $\boldsymbol{c}^{(m)}={c}_1^{(m)},\ldots,{c}_N^{(m)}$ according to Algorithm 2
\State Generate $\boldsymbol{\alpha}^{(m)}$  and $\boldsymbol{\beta}^{(m)}$ according to Algorithm 3
\EndWhile 
\Ensure $\boldsymbol{\alpha}^{(M)},\boldsymbol{\beta}^{(M)},\boldsymbol{c}^{(M)},
\boldsymbol{\gamma}^{(M)},\boldsymbol{\sigma^2}^{(M)},\boldsymbol{w}^{(M)}$

\end{algorithmic}
\end{algorithm}

\vskip 1cm
First, we describe how to sample from the conditional distribution 
$\pi(\boldsymbol{\gamma}|\boldsymbol{\alpha},\boldsymbol{\beta},\boldsymbol{c},\boldsymbol{\sigma^2},\boldsymbol{w},\boldsymbol{y})$,
 $\pi(\boldsymbol{\sigma^2}|
	\boldsymbol{\alpha},\boldsymbol{\beta},\boldsymbol{c},\boldsymbol{\gamma},\boldsymbol{w},\boldsymbol{y})$ and $\pi(\boldsymbol{w}|
	\boldsymbol{\alpha},\boldsymbol{\beta},\boldsymbol{c},\boldsymbol{\gamma},\boldsymbol{\sigma^2},\boldsymbol{y})$ since the conjugate priors of the parameters lead to the availability of closed forms of these conditional distributions (further details are reported in  Section \ref{sec:cond_dist} of the Supplementary Materials).
	
\subsubsection*{Conditional distribution of $\boldsymbol{\gamma}$}
We have
$
	\pi(\boldsymbol{\gamma}|\boldsymbol{\alpha},\boldsymbol{\beta},\boldsymbol{c},\boldsymbol{\sigma^2},\boldsymbol{w},\boldsymbol{y})=\prod_{k=1}^K \pi(\boldsymbol{\gamma}_k|{\sigma}^2_k,\boldsymbol{c},\boldsymbol{w},\boldsymbol{y})$,
where
$\pi(\boldsymbol{\gamma}_k|{\sigma}_k,\boldsymbol{c},\boldsymbol{w},\boldsymbol{y})
$	is  the  density of a multivariate Gaussian distribution  with vector mean 

$$
\boldsymbol{m}_{k}=\left(\frac{1}{g}\boldsymbol{I}+\sum_{i=1}^{N}\boldsymbol{\Psi}^\prime\boldsymbol{W}_{i} \boldsymbol{\Psi} I(c_i=k)\right)^{-1}\left(\sum_{i=1}^{N}\boldsymbol{\Psi}^\prime\boldsymbol{W}_{i} \mathbf{u}_{i}I(c_i=k)\right)
$$
and covariance matrix
$$
\boldsymbol{S}_k=\left(\frac{1}{g}\boldsymbol{I}+\sum_{i=1}^{N}\boldsymbol{\Psi}^\prime\boldsymbol{W}_{i} \boldsymbol{\Psi} I(c_i=k)\right)^{-1}.
$$
Here $\boldsymbol{W}_{i} $ is a diagonal matrix with entries $ [\sigma^2_{c_{i}}\omega^{2} w_{it}]^{-1} $, $ \boldsymbol{\Psi}=[\boldsymbol{\psi}(1),\ldots,\boldsymbol{\psi}(T)]^\prime $ is the matrix of covariates and $ \mathbf{u}_{i}=(u_{i1},...,u_{iT})' $, with $u_{it}=y_{it}-\tau\sigma_{c_i} w_{it}$.

\subsubsection*{Conditional distribution of $\boldsymbol{\sigma}^2$}

We note  that 
$
	\pi(\boldsymbol{\sigma^2}|
	\boldsymbol{\alpha},\boldsymbol{\beta},\boldsymbol{c},\boldsymbol{\gamma},\boldsymbol{w},\boldsymbol{y})
	=\prod_{k=1}^K
	\pi(\sigma_k^2|\boldsymbol{\gamma}_k,
	\boldsymbol{c},\boldsymbol{w},\boldsymbol{y})
$.
It turns out (see the Supplementary material) that $\pi(\sigma_k^2|\boldsymbol{\gamma}_k,
\boldsymbol{c},\boldsymbol{w},\boldsymbol{y})$
 is the distribution of an inverse Gamma random variable, $\operatorname{IG}(s_1,d_1)$, with  shape  parameter  and scale parameter
$$s_1=s_{0}+n_{k}T/2,\qquad 
d_1=d_{0}+\frac{1}{2\omega^{2}}\sum_{i=1}^{N}\left\{\sum_{t=1}^{T}\frac{\left(u_{it}-\boldsymbol{\psi}(t)^\prime \boldsymbol{\gamma}_{c_{i}}\right)^{2}}{w_{it}}\right\}I(c_i=k).$$

Random samples from the inverse Gamma distribution can be extracted from a Gamma distribution according to the relationship that  if $v \sim \mathrm{Gamma}(a, 1/b)$  then $1/v \sim \mathrm{InvGamma}(a, b)$.

\subsubsection*{Conditional distribution of $\boldsymbol{w}$}
We note  that 
$
	\pi(\boldsymbol{w}|
	\boldsymbol{\alpha},\boldsymbol{\beta},\boldsymbol{c},\boldsymbol{\gamma},\boldsymbol{\sigma^2},\boldsymbol{y})
	=\prod_{i=1}^{N}\prod_{t=1}^T
	\pi({w}_{it}|c_{i},\boldsymbol{\gamma}_{c_{i}},\sigma^2_{c_{i}},y_{it})
$. 
It turns out (see again the Supplementary Material) that the conditional distribution of  $w_{it}$ is  
the Generalized Inverse Gaussian distribution, $ \operatorname{GIG}(d,a,b), $ with  density
$$
	f(w;a,b,d)=
	\frac{(a/b)^{d/2}}{2 K_d(\sqrt{ab})} w^{(d-1)} e^{-(aw + b/w)/2},\qquad x>0,
$$	
where $K_d$ is a modified Bessel function of the second kind, $d=1/2$,
 $\displaystyle	a={(\tau^2+2\omega^{2})}/{\omega^2}$ and $\displaystyle b={(y_{it}-\boldsymbol{\psi}(t)^\prime\boldsymbol{\gamma}_{c_{i}})^{2}}/{(\sigma^2_{c_{i}}\omega^{2})}$.

We can use this trick to sample from this posterior distribution \citep{Jorgensen1982}. The density of the Inverse Gaussian distribution, $\operatorname{InvGIG}(\lambda,\mu)$,

\begin{eqnarray*}
	f(w;\lambda,\mu)&\propto& w^{-{3}/{2}}\exp\left\{\frac{-\lambda(w-\mu)^{2}}{2\mu^{2}w}\right\}\\
	&\propto& w^{-{3}/{2}}\exp\left\{-\frac{1}{2}\left(\frac{\lambda}{\mu^{2}}w+\frac{\lambda}{w}\right)\right\}
\end{eqnarray*}
is equal to the $\operatorname{GIG}(-1/2,{\lambda}/{\mu^{2}},\lambda)$.
We also know that if $v\sim\operatorname{GIG}(-1/2,b,a)$, then
$1/v\sim\operatorname{GIG}(1/2,a,b)$, i.e. 
if $v\sim\operatorname{InvG}({\lambda},\mu)$ then
$1/v\sim\operatorname{GIG}(1/2,\lambda,{\lambda}/{\mu^{2}})$.

Therefore, we can sample from the conditional posterior of $ w_{it} $ by drawing $v_{it}$ from  $\operatorname{InvG}({\lambda}/{\mu^{2}},\lambda)$ with
$\displaystyle\lambda=\frac{\tau^2+2\omega^{2}}{\omega^2}$ and $\displaystyle \frac{\lambda}{\mu^{2}}=
\frac{(y_{it}-\boldsymbol{\psi}(t)^\prime \boldsymbol{\gamma}_{c_{i}})^{2}}{\sigma^2_{c_{i}}\omega^{2}}$ that implies
$\displaystyle\mu=\frac{\sigma_{c_i}\sqrt{\tau^2+2\omega^{2}}}{|y_{it}-\boldsymbol{\psi}(t)^\prime \boldsymbol{\gamma}_{c_{i}}|}$ and considering $w_{it}=1/v_{it}$.

\subsubsection*{Conditional distribution of $\boldsymbol{c}$}

Generating $\boldsymbol{c}^{(m)}$ is difficult because of the normalizing constant $Z(\boldsymbol{\alpha},\boldsymbol{\beta})$.
     However, it is possible to embed another Gibbs step (see Algorithm \ref{eq:alg2}) into Algorithm 1 and construct an ergodic Markov chain whose stationary distribution is $\pi(\boldsymbol{\theta}|\boldsymbol{y})$.

\bigskip 


\begin{algorithm}
\caption{Gibbs step}\label{eq:alg2}
\begin{algorithmic}
\Require initial $\boldsymbol{c}^{m-1}$\\
1: $c_1^{m}\sim \pi(c_1|c_2^{m-1},\ldots, c_N^{m-1},\boldsymbol{\alpha},\boldsymbol{\beta},\boldsymbol{\gamma},\boldsymbol{\sigma^2},\boldsymbol{w},\boldsymbol{y})$\\
2: $c_2^{m}\sim \pi(c_2|c_1^{m},c_3^{m-1}\ldots, c_N^{m-1},\boldsymbol{\alpha},\boldsymbol{\beta},\boldsymbol{\gamma},\boldsymbol{\sigma^2},\boldsymbol{w},\boldsymbol{y})$\\
$\vdots$\\
N: $c_N^{m}\sim \pi(c_N|c_1^{m},\ldots, c_{N-1}^{m},\boldsymbol{\alpha},\boldsymbol{\beta},\boldsymbol{\gamma},\boldsymbol{\sigma^2},\boldsymbol{w},\boldsymbol{y})$
\Ensure $\boldsymbol{c}^{m}$

\end{algorithmic}
\end{algorithm}

\bigskip
\noindent The conditional distribution $\pi(c_i|c_j, j\ne i ,\boldsymbol{\alpha},\boldsymbol{\beta},\boldsymbol{\gamma},\boldsymbol{\sigma^2},\boldsymbol{w},\boldsymbol{y})$ is given by 
\paolo{
\begin{eqnarray*}
\pi(c_i|c_j, j\ne i ,\boldsymbol{\alpha},\boldsymbol{\beta},\boldsymbol{\gamma},\boldsymbol{\sigma^2},\boldsymbol{w},\boldsymbol{y})=
\cfrac{\exp\left\{-\frac 12\sum_{t=1}^T
\frac{(u_{it}-\boldsymbol{\psi}(t)^\prime\boldsymbol{\gamma}_{c_i})^{2}} 
{w_{it} \sigma^2_{c_i}\omega^2} +
	\alpha_{c_i} +
	\beta_{c_i} n_{c_i,i}\right\}}
{\sum_{k=1}^K 
	\exp\left\{-\frac 12\sum_{t=1}^T
 \frac{ (u_{it}-\boldsymbol{\psi}(t)^\prime\boldsymbol{\gamma}_{k})^{2}}{w_{it} \sigma^2_{k}\omega^2} +
 \alpha_{k} + 	\beta_k n_{k,i}\right\}},
\end{eqnarray*}
}
exploiting  equation \eqref{eq:condPotts}.

\subsubsection*{Conditional distribution of $\boldsymbol{\alpha}$ and $\boldsymbol{\beta}$}

The last step, i.e. the simulation from the conditional distributions of
$\boldsymbol{\alpha}^{(m)}$ and $\boldsymbol{\beta}^{(m)}$ is performed using a Metropolis-Hastings algorithm. 
However, the direct implementation of the algorithm is still difficult due to the presence of the normalizing constant $Z(\boldsymbol{\alpha},\boldsymbol{\beta})$.

\citet{marjoram2003markov} have shown that it is possible to define a valid likelihood-free MH algorithm for posterior distributions with intractable normalizing constants by introducing a carefully selected auxiliary variable and a tractable sufficient statistic on the target density.

Since exact likelihood-free MH algorithms have several shortcomings described in \citet{marin2012approximate}, we modify the algorithm described in \citet{Pereyra:2013} that leads to an Approximate Bayesian Computation (ABC) with a probability-free Metropolis-Hastings step. The considered algorithm is described in the next box (Algorithm \ref{eq:alg3}).

\begin{algorithm}
\caption{ABC likelihood-free MH step for $\boldsymbol{\alpha}$ and $\boldsymbol{\beta}$.} \label{eq:alg3}
\begin{algorithmic}[1]
\Require initial $\boldsymbol{\alpha}^{m-1},\boldsymbol{\beta}^{m-1}$
\State Fix $\alpha_1^*=0$ 
\State Generate proposals $\alpha^*_k\sim\mathcal{N}(\alpha_k^{m-1},\sigma^2_\alpha)$ for $k=2,\ldots,K$ 
\State Generate $\beta_k^{*} \sim \mathcal{NT}_{0,B}(\beta_k^{m-1},\sigma^2_{\beta})$\ for $k=1,\ldots,K$
\State Generate an auxiliary $\boldsymbol{c}^*$ from the distribution	$f(\boldsymbol{c}|\boldsymbol{\alpha}^*,\boldsymbol{\beta}^{*})$ in \eqref{eq:Potts}\;
\If{$\rho(\boldsymbol{\eta}(\boldsymbol{c}^*),\boldsymbol{\eta}(\boldsymbol{c}^{m-1}))  < \mbox{tolerance} $} 
\State Set $\displaystyle r=\frac{\pi(\boldsymbol{\alpha}^{*})\pi(\boldsymbol{\beta^{*}})\,
			\prod_{k=2}^K\phi(\alpha_k^{m-1}|\alpha_k^*,\sigma^2_\alpha)
   \prod_{k=1}^K\phi_{[0,5]} (\beta^{m-1}_k|\beta^*_k,\sigma^2_\beta)
			}{\pi(\boldsymbol{\alpha}^{m-1})\pi(\boldsymbol{\beta}^{m-1})\, \prod_{k=2}^K\phi(\alpha_k^{*}|\alpha_k^{m-1},\sigma^2_\alpha)\prod_{k=1}^K\phi_{[0,5]}(\beta^*_k|\beta^{m-1}_k,\sigma^2_\beta)}$\;
\State 	Draw $u \sim \mathcal{U}(0,1)$\;  
\If{$r < u$}
\State $\boldsymbol{\alpha}^{m}=\boldsymbol{\alpha}^{*},\boldsymbol{\beta}^{m}=\boldsymbol{\beta^{*}}$\;
\Else 
\State $\boldsymbol{\alpha}^{m}=\boldsymbol{\alpha}^{m-1},\boldsymbol{\beta}^{m}=\boldsymbol{\beta}^{m-1}$\;
\EndIf
\Else
\State $\boldsymbol{\alpha}^{m}=\boldsymbol{\alpha}^{m-1},\boldsymbol{\beta}^{m}=\boldsymbol{\beta}^{m-1}$\;
\EndIf
\Ensure $\boldsymbol{\alpha}^{m},\boldsymbol{\beta}^{m}$

\end{algorithmic}
\end{algorithm}

\bigskip
\noindent \paolo{Here $\mathcal{NT}_{a,b}(\mu,\sigma^2)$ is the truncated Gaussian distribution with mean $\mu$ and variance $\sigma^2$, within the interval $( a , b )$, with $-\infty \leq a<b\leq +\infty$. 
Here, the vector $\boldsymbol{\eta}(\cdot)$ serves as a vector of statistics, while the function $\rho(\cdot)$ functions as a distance metric.}

\paolo{The entire MCMC algorithm necessitates the specification of initial values in Algorithm 1 and variances for the proposals, the vector $\boldsymbol{\eta}(\cdot)$,  the distance $\rho(\cdot)$  and the tolerance in Algorithm 3.}

\paolo{In all numerical experiments, the initial values  $\boldsymbol{w}^{(0)}$ were randomly assigned.
The algorithm outlined in \citet{Sottile:2019} yields the memberships $ \boldsymbol{c}^{(0)}$, while the values of $\boldsymbol{\alpha}^{(0)},\boldsymbol{\gamma}^{(0)},\boldsymbol{\sigma^2}^{(0)}$ are estimated by a multinomial regression model ($\boldsymbol{\alpha}^{(0)}$) and K different quantile regressions ($\boldsymbol{\gamma}^{(0)},\boldsymbol{\sigma^2}^{(0)}$), respectively. 
The initial value for the spatial dependence parameter differs in the simulations and in the real data application. Specifically, we set ${\beta}_k^{(0)}$ equal to 2 for $k=1,\ldots,K$ in the simulations and 10 in the real data application. This choice was arbitrary and motivated solely by the sensitivity of the sampler's mixing properties to the starting values, resulting in prolonged burn-in periods.}

\paolo{We set the variance of proposals $\sigma^2_{\alpha}=0.1$ for $\boldsymbol{\alpha}$. The variance of proposals $\sigma^2_{\beta}$ depends on the choice of the uniform $\mathcal{U}(0,B)$ as prior for $\boldsymbol{\beta}$.  Since we set $B=10$ for the simulation experiments and 
$B=100$ for the real data example, we specify
$\sigma^2_{\beta}$ as 0.04 and 1, respectively. We select $\boldsymbol{\eta}(\boldsymbol{c}) 
=(\eta_1(\boldsymbol{c}),\ldots,\eta_{2K-1}(\boldsymbol{c}))^\prime
=\left(n_2,\ldots,n_K, \sum_{i=1}^N n_{1,i},\ldots,\sum_{i=1}^N n_{K,i}\right)^\prime,$ since the vector represents sufficient statistics in the Potts model \eqref{eq:Potts}.
The function
$\rho(\boldsymbol{\eta}(\boldsymbol{c}^*),\boldsymbol{\eta}(\boldsymbol{c}^{m-1}))=\frac{1}{2K-1}\sum_{j=1}^{2K-1}\frac{\left|{\eta}_j(\boldsymbol{c}^*)-{\eta}_j(\boldsymbol{c}^{m-1})\right|}{\eta_j(\boldsymbol{c}^{m-1})},$
measures the distance between $\boldsymbol{\eta}(\boldsymbol{c}^*)$ and $\boldsymbol{\eta}(\boldsymbol{c}^{m-1})$. }

\paolo{The tolerance is adjusted to achieve a 5\% acceptance rate. This value is calculated using the 5th percentile of the observed tolerances $\rho(\boldsymbol{\eta}(\boldsymbol{c}^*),\boldsymbol{\eta}(\boldsymbol{c}^{m-1}))$ during the burning phase.}


\section{Simulations}\label{sec4}
\begin{figure}[ht]
\centering
\includegraphics[width=0.49\linewidth]{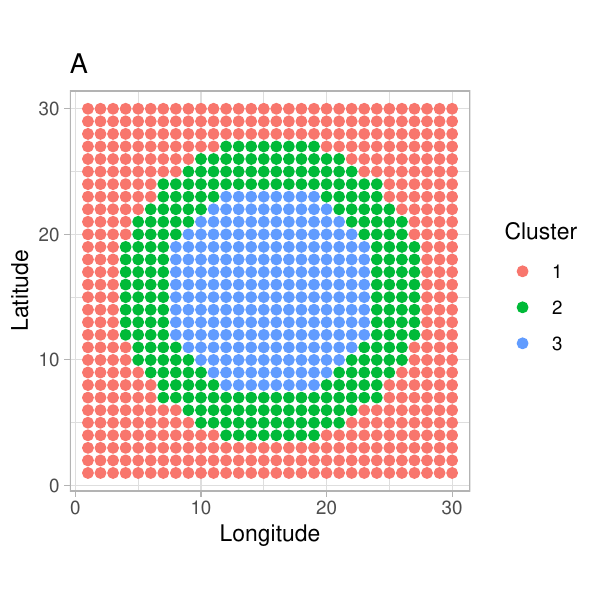}
\includegraphics[width=0.49\linewidth]{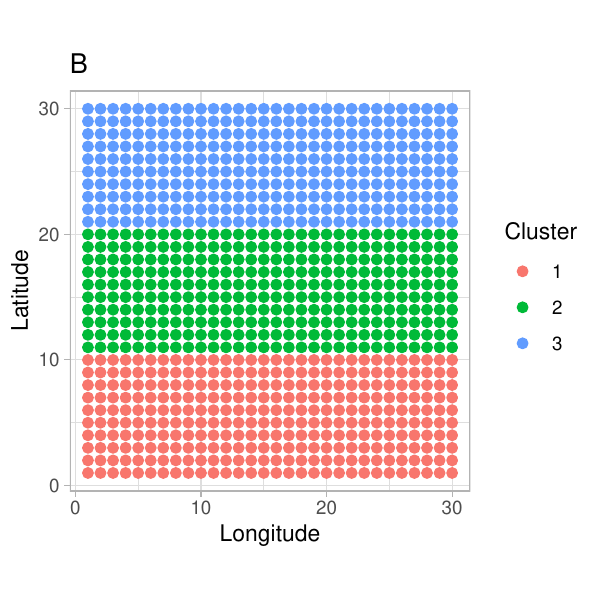}
\caption{Spatial pattern of concentric circles (A) and flaglike rectangles (B). }\label{fig:4}
\end{figure}

\noindent
\paolo{We conducted simulations to illustrate how our spatial clustering method can distinguish three different groups of time series, taking into account both the temporal pattern and the marginal error distribution.}  Two membership allocations (concentric circles and flaglike rectangles) of $N=900$ sites on a  $30\times 30$ regular grid will be tested. We consider three groups for each allocation, as in Figure \ref{fig:4}.  Concentric circles have different group sizes, namely $440$, $244$, and $216$. The flaglike rectangle is divided into groups of the same size.

\subsection{First simulation example}
\begin{figure}[ht]
	\centering
	\includegraphics[width=0.8\linewidth]{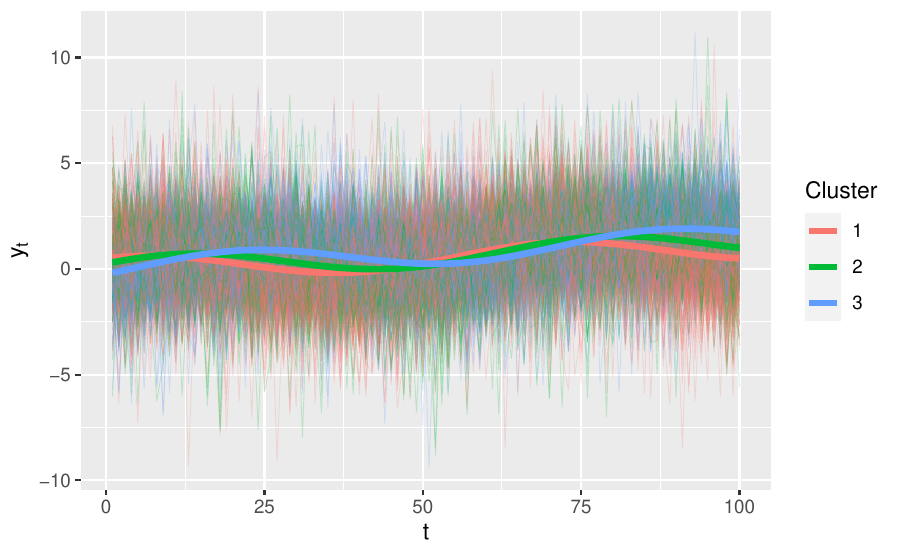}
	\caption{Simulated time series with ALD random error. The mean trend for each cluster is reported with a solid line.} \label{fig:5}
\end{figure}

\paolo{The first simulation example attempts to measure the accuracy of clustering and parameter estimation when in the joint distribution of a time series is \eqref{eq:fy}  a spatial pattern for membership is included.
For each site in the group $k$ 
 \typo{(with $k=1,2,3$)} a time series of $T=100$ observations is simulated based on the model
\begin{equation}\label{eq:modsimulation}
y_{t,k}=\mu_k(t) + e_{t,k}, \qquad t=1,\ldots,T. 
\end{equation}
The function
\begin{equation}\label{eq:mu}
\mu_k(t)= \frac{\theta_{0k} }{100}t+  \theta_{1k}  \cos\left(\frac{3\pi }{100}t\right) +\theta_{2k} \sin\left(\frac{3\pi }{100}t\right)   
\end{equation} represents an overall trend and a cyclic pattern.
We set the parameters for three clusters as follows: $(\theta_{01},\theta_{11},\theta_{21})=(1,0.5,0.25)$,
$(\theta_{02},\theta_{12},\theta_{22})=(1.25,0.25,0.5)$ and 
$(\theta_{03},\theta_{13},\theta_{23})=(1.5,-0.25,0.5)$, and it is assumed that $e_{t,k}\sim \operatorname{ALD}(p,\sigma)$ with $p=0.5$ and $\sigma=0.5$. It is worth mentioning that the presence of random ALD errors complicates the time series clustering situation, as shown in Figure \ref{fig:5}.
The simulation experiment is repeated $100$ times.}

\paolo{We evaluated the effectiveness of two clustering techniques: Asymmetric Laplace Model-based Clustering with Spatial Dependence (ALC-S) and the spatially-independent procedure (ALC), where $\beta_k=0$ for $k=1,\ldots,K$ in equation \eqref{eq:Potts}.}

\paolo{In the estimation stage the vector $\mathbf{\psi}(t)$ in \eqref{eq:regression} is 
 $(t/100,\cos(3\pi\,
t/100), \sin(3\pi\, t/100))^{\prime}$ to facilitate a comparison between the estimated and actual parameters. 
We use the MCMC algorithm for 1000
iterations to obtain posterior estimates, discarding the first $300$ as burn-in. Based on trace plots, this number of iterations appears to be adequate for Markov chain convergence. The group membership is determined by the posterior mode of $c_i$, while the regression parameter estimates are calculated using the posterior mean.
For the ALC-S method, three nearest neighbor networks have been considered: specifically, the 4-NN, 8-NN, and 12-NN. }

\paolo{
The performance of both clustering algorithms in \typo{reconstructing} the three clusters is evaluated by gauging the degree of agreement between the estimated partition and the actual partition, as determined by the Adjusted Rand Index (ARI) \citep{Hubert1985}. The results are outlined in Table \ref{simARI1}. 
The \typo{results} suggest that the \typo{inclusion} of the spatial layer \typo{improves} the consistency between the clustering and the true spatial arrangement. On average, this enhancement leads to a 13-14\% improvement in \typo{the} ARI values for both the 4-NN spatial configurations and results in values greater than 90\%. The concordance in the concentric circle scenario decreases as the number of neighbors increases. As reported in Table \ref{simtheta1}, both methods \typo{show} exceptional performance in estimating the actual curve parameters. \typo{In particular}, ALC-S exhibits higher accuracy, although it may be slightly less \typo{accurate} in certain situations such as the concentric circle scenario.
}
\begin{table}[ht]
\caption{\label{simARI1} First simulation example. Average ARI values of $100$ replications for each experiment and clustering method (standard error between parentheses).} 
\begin{tabular}{lcc}
\toprule
& \textbf{Concentric circles} & \textbf{Flaglike rectangles}\\
\hline
\textbf{ALC-S with 4-NN} &  $ 0.914\, (0.062) $ & $ 0.923\, (0.075)  $\\
\textbf{ALC-S with 8-NN} &  $ 0.861\, (0.157) $ & $ 0.934\, (0.122)  $\\
\textbf{ALC-S with 12-NN} &  $ 0.801\, (0.194) $ & $ 0.941\, (0.145)  $\\
\textbf{ALC} &    $ 0.794\, (0.043) $ & $ 0.762\, (0.028) $\\
\bottomrule
\end{tabular}
\end{table}
\paolo{
\begin{table}[ht]
\caption{\label{simtheta1} First simulation example. Average estimates for $\theta$ were obtained from 100 replications for each clustering method. Standard errors are reported in parentheses.} 
\begin{tabular}{lccc}
\toprule
\multicolumn{3}{l}{\textbf{Concentric circles}} & \\
& \multicolumn{3}{c}{\textbf{Cluster 1}}\\
&$\theta_{10} = 1$&$\theta_{11} = 0.5 $&$\theta_{12}= 0.25$\\
\hline
\textbf{ALC-S with 4-NN} & 1.001 (0.012) & 0.504 (0.011)& 0.244 (0.015)    \\
\textbf{ALC-S with 8-NN} & 1.002 (0.016) & 0.500 (0.012)& 0.248 (0.018)    \\
\textbf{ALC-S with 12-NN} & 1.026 (0.035)& 0.475 (0.046)   & 0.271 (0.035) \\
\textbf{ALC}   & 1.001 (0.010)  & 0.508 (0.010) & 0.240 (0.013)   \\
&\multicolumn{3}{c}{\textbf{Cluster 2}}\\
&$\theta_{20}= 1.25 $&$\theta_{21} = 0.25 $&$\theta_{22} = 0.5 $\\
\hline
\textbf{ALC-S with 4-NN}& 1.252 (0.012) & 0.239 (0.023)& 0.510 (0.017)   \\
\textbf{ALC-S with 8-NN}& 1.253 (0.024) & 0.240 (0.029) & 0.510 (0.025)  \\
\textbf{ALC-S with 12-NN}& 1.256 (0.066)& 0.224 (0.097) & 0.509 (0.070)  \\
\textbf{ALC} & 1.244 (0.021) & 0.237 (0.021)& 0.504 (0.015)    \\
& \multicolumn{3}{c}{\textbf{Cluster 3}}\\
& $\theta_{30} = 1.5 $&$\theta_{31} = -0.25 $&$\theta_{32}= 0.5$\\
\hline
\textbf{ALC-S with 4-NN}& 1.502 (0.014) & -0.256 (0.017) & 0.497 (0.010)   \\
\textbf{ALC-S with 8-NN}& 1.500 (0.011) & -0.253 (0.016)& 0.500 (0.013)    \\
\textbf{ALC-S with 12-NN}& 1.481 (0.047)& -0.212 (0.078) & 0.497 (0.023)   \\
\textbf{ALC}  & 1.500 (0.012)  & -0.256 (0.015)& 0.500 (0.011)   \\
\bottomrule
&&&\\
\hline
\multicolumn{3}{l}{\textbf{Flaglike rectangles}} & \\
& \multicolumn{3}{c}{\textbf{Cluster 1}}\\
&$\theta_{10} = 1$&$\theta_{11} = 0.5 $&$\theta_{12}= 0.25$\\
\hline
\textbf{ALC-S with 4-NN} & 0.995 (0.017) & 0.511 (0.020)& 0.237 (0.021)    \\
\textbf{ALC-S with 8-NN} & 0.999 (0.013) & 0.503 (0.009)& 0.246 (0.009)    \\
\textbf{ALC-S with 12-NN} & 1.015 (0.043)& 0.487 (0.043) & 0.264 (0.042)   \\
\textbf{ALC}  & 1.005 (0.018) & 0.507 (0.018)& 0.242 (0.017)    \\
&\multicolumn{3}{c}{\textbf{Cluster 2}}\\
&$\theta_{20}= 1.25 $&$\theta_{21} = 0.25 $&$\theta_{22} = 0.5 $\\
\hline
\textbf{ALC-S with 4-NN} & 1.247 (0.021) & 0.248 (0.025)& 0.501 (0.021)    \\
\textbf{ALC-S with 8-NN} & 1.250 (0.011) & 0.249 (0.009)& 0.503 (0.009)    \\
\textbf{ALC-S with 12-NN} & 1.251 (0.038)& 0.244 (0.031)  & 0.510 (0.057)  \\
\textbf{ALC}  & 1.251 (0.022) & 0.237 (0.028)& 0.504 (0.018)    \\
& \multicolumn{3}{c}{\textbf{Cluster 3}}\\
& $\theta_{30} = 1.5 $&$\theta_{31} = -0.25 $&$\theta_{32}= -0.5$\\
\hline
\textbf{ALC-S with 4-NN} & 1.501 (0.012) & -0.257 (0.013)& 0.501 (0.011)   \\
\textbf{ALC-S with 8-NN} & 1.498 (0.010) & -0.253 (0.009)& 0.501 (0.008)   \\
\textbf{ALC-S with 12-NN} & 1.502 (0.011)& -0.252 (0.009) & 0.501 (0.011)  \\
\textbf{ALC} & 1.496 (0.012) & -0.255 (0.019)& 0.500 (0.009)    \\
\bottomrule
\end{tabular}
\end{table}
}

\subsection{Second simulation example}
\paolo{
The second, more extensive example compares the proposed procedure with \typo{nice competing} procedures that have been implemented in R. The \typo{goal} is still to provide a comprehensive analysis and evaluation of the effectiveness of the proposed procedure relative to the alternatives.
We \typo{keep} the two spatial patterns previously analyzed, the same number of time series observations, and the same functions \eqref{eq:mu} for the clusters. 
}

\paolo{
For the distribution of the error term in \eqref{eq:modsimulation}, we defined two different scenarios (A and B). In scenario A, it is assumed that $e_{t,k}$ follows a Gaussian distribution with mean 0 and standard deviation 1. This scenario is ideal for model-based Gaussian clustering to achieve high efficacy.}

\begin{figure}[ht]
\begin{center}
\begin{tabular}{c}
	\includegraphics[width=0.7\linewidth]{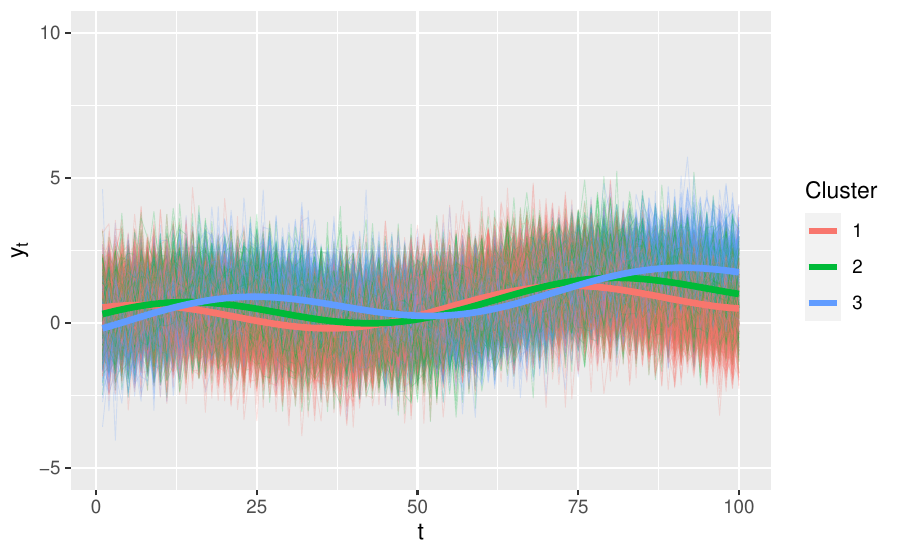}
 \\
 \includegraphics[width=0.7\linewidth]{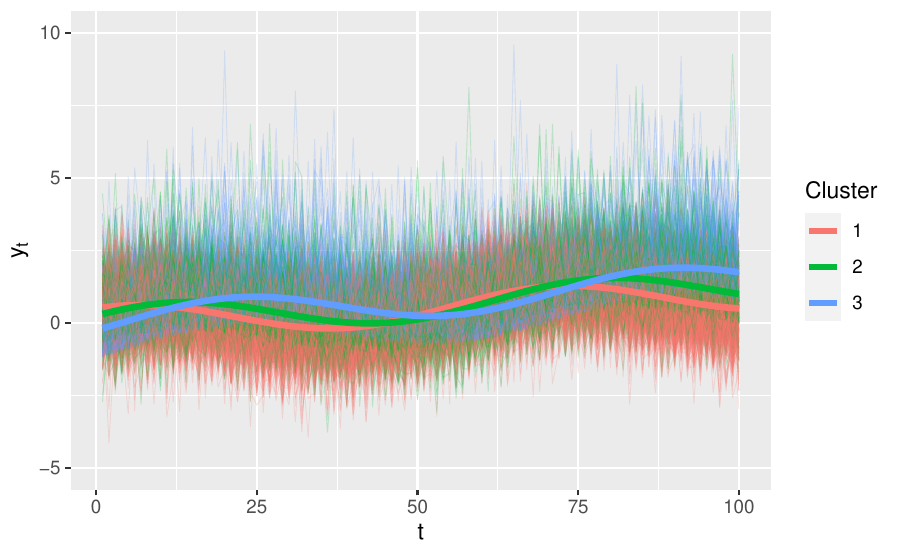}
\end{tabular}
\end{center}
	\caption{Simulated time series under scenario A  (top) and B (bottom).} \label{fig:6}
\end{figure}

\paolo{
In scenario B, for each cluster we define the following error distribution for 
\begin{enumerate}
\item $e_{t,1}=N_t$;
\item $e_{t,2} = U_t N_t +(1-U_t)(G_t-1)$;
\item $e_{t,3}=G_t-1 $,
\end{enumerate}
 with $U_t \sim Ber(0.5)$, $N_t\sim \mathcal{N}(0,1)$  and $G_t\sim\mbox{Gamma}(1,1)$ independent random variables.
The three error distributions share the same mean of $0$ and standard deviation of $1$. However, the skewness \typo{of} the error distribution increases with the \typo{cluster number}.
}
\paolo{
An example of a simulated series for both scenarios is reported in Figure \ref{fig:6}.
}
\paolo{We estimate the function $\mu_k(t)$ for each cluster using a $J$ cubic B-spline basis in \eqref{eq:regression}, with equally spaced knots over $[0,T]$. 
The preliminary selection of $J$ is done using a BIC-like overall criterion as follows.
}
\paolo{
\begin{enumerate}
\item[1.] We estimate the parameter  ${\boldsymbol{\gamma}}_i$ of model \eqref{eq:fy} for each time series ${\boldsymbol{y}}_i$ by minimizing $\sum_{t=1}^{T}\rho_{p}(y_{i,t}-\boldsymbol{\psi}(t)^\prime\boldsymbol{\gamma}_i)$.
\item[2.] We evaluate the BIC-like criterion
$
BIC_i(J)=\sum_{t=1}^{T}\rho_{p}(y_{it}-\boldsymbol{\psi}(t)^\prime
\widehat{\boldsymbol{\gamma}}_i)+J \log(T)
$.
\item[3.] We  find the value $J$ that minimizes the overall  value
$\overline{BIC}(J)=\sum_{i=1}^N BIC_i(J)$.
\end{enumerate}
}
\paolo{
The minimum value was achieved at $J=6$ in the experiment. We continue to utilize the MCMC algorithm for estimation, completing $1000$ iterations (with $300$ as burn-in) for the ALC-S procedure utilizing an 8-NN network.
\typo{Nine} clustering methods were assessed as competitors.}
\paolo{
\begin{enumerate}
	\item [1)] Hierarchical Clustering based on the raw data (HC-R). We use the DIvisive ANAlysis Clustering (DIANA) algorithm \citep{kaufman1990rousseeuw}, 
	with the dissimilarity measure between two time series $\boldsymbol{y}_{i}$ and $\boldsymbol{y}_{i'}$  $d_{ij}^R=\|\boldsymbol{y}_{i}-\boldsymbol{y}_j \|$ 
	and we extract three clusters.
	\item [2)] Hierarchical Clustering based on the Least Squares regression estimates  (HC-LS).  For each time series, we get the least squares estimates of 	 ${\boldsymbol{\gamma}}_i$ in \eqref{eq:regression},
	 $\widehat{\boldsymbol{\gamma}}^{LS}_i$,
	 and  we use the DIANA algorithm with  the  dissimilarity measure  $d_{ij}^{LS}=\|\boldsymbol{\gamma}^{LS}_{i}-\boldsymbol{\gamma}^{LS}_{j}\|$ 
	\item [3)] Hierarchical Clustering based on the Least Absolute Deviation regression estimates (HC-AD): we get the absolute deviation regression estimates of ${\boldsymbol{\gamma}}_i$ in \eqref{eq:regression}, $\widehat{\boldsymbol{\gamma}}^{AD}_i$ and we still use the DIANA algorithm with $d_{ij}^{AD}=\|\boldsymbol{\gamma}^{AD}_{i}-\boldsymbol{\gamma}^{AD}_{j}\|$. 
\paolo{\item [4)] Geostatistical Spectral Clustering (GSC) which is based on the spectral decomposition of the graph Laplacian \citep{romary:2015};
\item [5)] Functional Principal Components Analysis Clustering (FPCAC) based on the principal component analysis for functional data obtained by quantile regression \citep{Sottile:2019} as implemented in the  R package \texttt{clustEff};
\item [6)] Functional Data Clustering Using Adaptive Density Peak Detection (FADP) which the density is estimated by functional k-nearest neighbor density estimation based on a proposed semi-metric between functions \citep{wang2017fast} as implemented in the  R package \texttt{FADPclust};}
\typo{\item [7)] The FunEM algorithm is based on a discriminative functional mixture model that enables data clustering  in a unique and discriminative functional subspace \citep{Bouveyron:2015}. The procedure is implemented using the R package \texttt{funFEM};
\item [8)]  Functional data clustering method for High-Dimensional Data Clustering (FunHDDC), based on a functional latent mixture model \citep{Bouveyron:2011}. The algorithm is implemented in the R package \texttt{funHDDC};}
 \item [\typo{9})] Asymmetric Laplace model-based Clustering without spatial dependence (ALC).
\end{enumerate}
}
\paolo{
To address possible spatial dependence, we modify the measure of dissimilarity between competitors 1-3.  Specifically, we define $d^w_{ij}$ as $d_{ij}$ multiplied by $\gamma(h_{ij})$, where $\gamma(h_{i,j})$ is the variogram computed for the distance $h_{i,j}$ between site $i$ and site $j$. We consider an exponential model with a nugget effect for the variogram. We fit the variogram using regression scores derived from the first principal component extracted from the raw data \citep{Oliver:1989}.
}
\paolo{\begin{table}[ht]
\caption{\label{simARI2}Average ARI values of 100 replications for each experiment and clustering method (standard error between parentheses).} 
\begin{tabular}{lcccc}
\toprule
& \multicolumn{2}{c}{\textbf{Concentric circles}} & \multicolumn{2}{c}{\textbf{Flaglike rectangles}}\\
&Scenario A & Scenario B & Scenario A & Scenario B \\
\hline
\textbf{HC-R} &  $ 0.467\, (0.111) $ & $ 0.517\, (0.103) $&
$ 0.736\, (0.176) $& $ 0.488\, (0.148) $\\
\textbf{HC-LS} &  $ 0.392\, (0.074) $ & $ 0.392\, (0.076) $
& $ 0.421\, (0.040) $& $ 0.418\, (0.045) $\\
\textbf{HC-AD}  &  $ 0.318\, (0.074) $ & $ 0.261\, (0.056) $
& $ 0.322\, (0.032) $& $ 0.335\, (0.037) $\\
\textbf{GSC}   &  $ 0.348\, (0.067) $ & $ 0.537\, (0.351) $
& $ 0.459\, (0.181) $& $ 0.559\, (0.417) $\\
\textbf{FPCAC}  & $ 0.828\, (0.020) $ & $ 0.826\, (0.026) $
& $ 0.838\, (0.021)  $& $ 0.844\, (0.024) $\\
\textbf{FADP}  & $ 0.730\, (0.062) $ & $ 0.667\, (0.131) $
& $ 0.728\, (0.065) $& $ 0.728\, (0.072) $\\
\textbf{FunFEM}  & $ 0.169\, (0.020) $ & $ 0.172\, (0.017) $
& $ 0.179\, (0.024)  $& $ 0.168\, (0.019) $\\
\textbf{FunHDDC}  & $ 0.370\, (0.082) $ & $ 0.418\, (0.062) $
& $ 0.421\, (0.025)  $& $ 0.412\, (0.052) $\\
\textbf{ALC} &  $ 0.752\, (0.038) $ & $ 0.703\, (0.023) $
& $ 0.717\, (0.038) $& $ 0.685\, (0.032) $\\
\textbf{ALC-S} &  $ 0.815\, (0.066) $ & $ 0.866\, (0.135) $
& $ 0.808\, (0.077) $& $ 0.968\, (0.043) $\\
\bottomrule
\end{tabular}
\end{table}
}

\paolo{
We evaluated the performance of clustering utilizing the adjusted Rand index (ARI) \citep{Hubert1985}. 
Table \ref{simARI2} displays the adjusted Rand Index (ARI) values for the \typo{ten} clustering approaches in each scenario. From these findings, it is evident that our clustering algorithm demonstrated an overall satisfactory performance (ARI $> 0.80$). In contrast, distance matrix-based methods (competitors 1-3) showed low levels of agreement, especially in the presence of a complex spatial pattern (concentric circles). A similar behavior was observed in the GSC procedure. Non-spatial methods based on robust quantile regression (FPCAC and FADP) worked moderately well, \typo{while the two considered procedures for functional data clustering (FunFEM and FunHDDC) reported poor classification performances.}
Taking into account scenario B, ALC-S reported the highest ARI values (0.866 and 0.968, for concentric circles and flag-like rectangles, respectively). Our proposal exhibited an increase in the average classification accuracy in each scenario upon the addition of spatial information, as previously stated in the previous simulation subsection.}

\section{Clustering of the Mediterranean Sea SST}\label{sec5}

In this section, we perform a spatial clustering of the Mediterranean SST monthly time series, considering a temporal window between 1982 and 2012; the reference data are previously presented in Section \ref{sec2}. As shown in Figure \ref{fig:2}, we can observe a strong and time-varying seasonal pattern, \paolo{i.e. non-stationarity in time, with the presence of a maximum in late summer and a minimum in the spring months.
Another distinctive element of this dataset is the dependence of the mean sea surface temperature on latitude.} Higher average values are observed \paolo{at} sites close to the African continent, while near the Gulf of Lion we can \paolo{find} the coldest water masses. 
\paolo{ To eliminate the effect of latitude/longitude, we convert each time series  to its normalized anomalies 
$(y_{it}-\bar{y}_i)/s_{i}$, $t=1,\ldots,T$.
Here, $\bar{y}_i$ and $s_{i}$ represent the sample mean and standard deviation, respectively, of the time series $i$.
After the transformation, the Potts model specified in equation \eqref{eq:Potts} seems appropriate for modeling spatial membership clustering.
The network specification based on sites and their neighbors allows the consideration of non-convex domains. To account for edge effects, the number of neighbors of a site at the edge has been modified.
\typo{Apart from edge effects, the spatial dependence between clusters} is assumed to be different for each group and isotropic. This fact may be a limitation, but a parameterization that would allow for anisotropy would require a large number of parameters. 
}

We specify a seasonal modulation model \citep{eilers2008modulation}, namely
 \begin{equation*}
    \mu(t)=m(t)+ s(t) 
 \end{equation*}
 where the component $m(t)$ defines an overall trend while the seasonality $s(t)$ can be expressed as the sum of $D$ harmonics
  \begin{equation*}
   s(t)= \sum_{d=1}^D \left\{ g_{2d}(t) \cos(\pi d t/6)+ g_{3d}(t) \sin(\pi d t/6)\right\}.
 \end{equation*}
In our example, we consider $D=2$, \typo{to capture}  intra-annual variations in the seasonal component.
The model can be structured by a linear combination of $J_1$, $2 \times J_2$, and $2 \times J_3$ cubic B-splines, with equally spaced knots over the time interval $[1,372]$.
It is easy to see that the resulting $\mu(t)$, namely 
 \begin{eqnarray}
	\mu(t)&=&\sum_{j=1}^{J_1} \gamma_{1l} \psi_{1j}(t)+ \sum_{d=1}^2   \left\{ \left[\sum_{j=1}^{J_2} \gamma_{2dj}\psi_{2dj}(t)\right]\cos(\pi d t/6)+ \left[\sum_{j=1}^{J_3} \gamma_{3dj}\psi_{3dj}(t)\right]\sin(\pi d t/6)\right\}\nonumber\\
 &&\label{eq:harmonic},
\end{eqnarray}
can be written as a linear combination of known functions. 

\paolo{A preliminary data analysis was performed to determine the number of
functions in \eqref{eq:harmonic}} and the number of clusters. Following the strategy outlined in Section \ref{sec4}, we minimize the total $\operatorname{BIC}(J_1,J_2,J_3)$ using a median (i.e. $p=0.5$) regression, varying the value of $J_1$, $J_2$, and $J_3$ from 4 to 8.

The \paolo{total value} $\operatorname{BIC}(J_1,J_2,J_3)$ is reported in the Supplementary Materials (Table \ref{tab:BIC}), where the minimum value is reported for $J_1=4$ for the trend (including an internal intercept)\paolo{, $J_2=8$, and $J_3=4$}. As a neighborhood network for Potts model \ref{eq:Potts}, we have considered the {4-NN  and 8-NN network}.

Then, we identify the number of clusters that minimize the Widely Applicable Information Criterion ($\operatorname{WAIC}$)  \citep{watanabe:2010}, computed as follows:
$$\operatorname{WAIC} = -2\left(\sum_{i=1}^N \log \mathbb{E}_{\boldsymbol{\theta},c_i}[f(\boldsymbol{y}_i|\boldsymbol{\theta},c_i)|\boldsymbol{y}]-\sum_{i=1}^N  \mathbb{V}_{\boldsymbol{\theta},c_i} [ \log f(\boldsymbol{y}_i|\boldsymbol{\theta},c_i)|\boldsymbol{y}]\right). $$
The expectation $\mathbb{E}_{\boldsymbol{\theta},c_i}[f(\boldsymbol{y}_i|\boldsymbol{\theta},c_i)|\boldsymbol{y}]$ and the variance
$\mathbb{V}_{\boldsymbol{\theta},c_i} [ \log f(\boldsymbol{y}_i|\boldsymbol{\theta},c_i)|\boldsymbol{y}]$ are  estimated by the sample mean of $ f(\boldsymbol{y}_i|\boldsymbol{\theta}^{(m)},c_i^{(m)})$, and the sample
variance of $\log f(\boldsymbol{y}_i|\boldsymbol{\theta}^{(m)},c_i^{(m)})$, $m=1\ldots,M$ of the posterior simulations derived from running the MCMC algorithm after convergence.

\paolo{The WAIC index is calculated for a range of clusters, specifically from 2 to 8. As shown in the Supplementary Materials, we found this minimum to be equal to 8 and 6 clusters for the 4-NN and 8-NN networks, respectively.} The final membership ${c}_i$ and coefficients $\boldsymbol{\gamma_k}$ were estimated using the mode and mean, respectively, of the posterior distribution after 3000 MCMC iterations, with the first 500 iterations discarded as burn-in. 

\paolo{
We present the clustering results obtained from grouping data into $K=6$ clusters with the 8-NN network and compare them to the results obtained using the ALC method, disregarding spatial dependence. 
}
\begin{figure}[ht!]
  \centering
\begin{tabular}{c}
\includegraphics[width=0.99\linewidth]{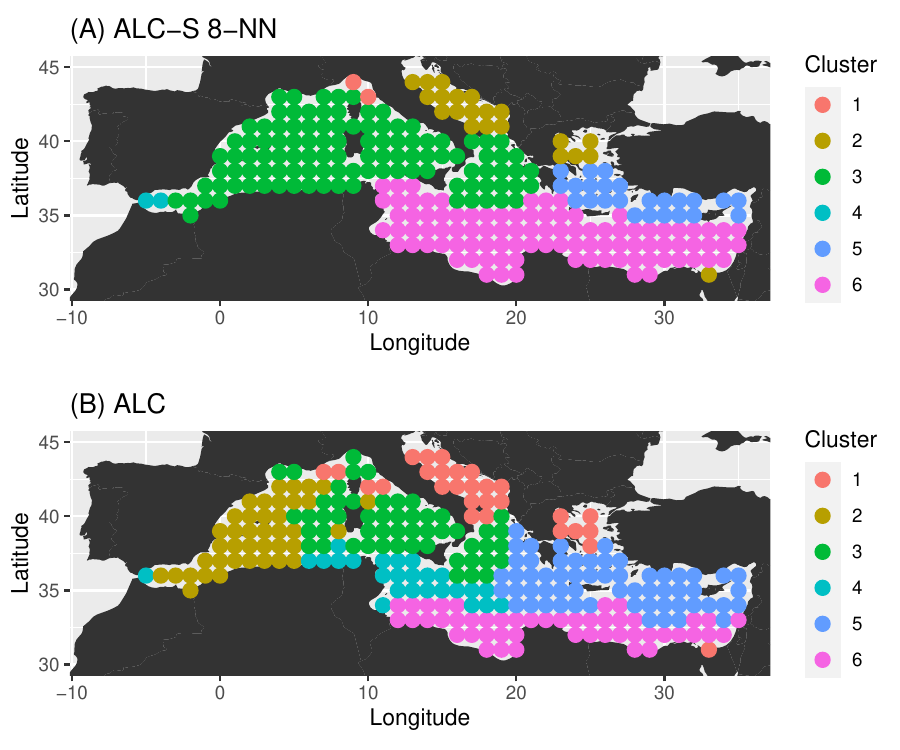}
\end{tabular}
	\caption{Spatial clustering by the ALC and ALC-S for K=6 clusters\label{figure_comparison1}. }
\end{figure}

\paolo{Cluster 1 characterizes a small area that overlaps with the Gulf of Genoa, whereas Cluster 2 in dark yellow encompasses the shallow waters of the Adriatic Sea, the western Aegean Sea, and the Nile's mouths. Cluster 3 extends to the central-western part of the Mediterranean Sea, while 
Cluster 4 in light blue defines another small area that coincides with the Strait of Gibraltar. Cluster 5, depicted in dark blue, encompasses the portion of the Mediterranean Sea that lies between the coasts of Turkey and those of Greece, as well as Cyprus and the island of Crete. Cluster 6 denotes a distinct area within the central-southern Mediterranean Sea, extending from Crete to the Egyptian coasts. 
Compared to the ALC procedure, the ALC-S \typo{provides} a more consistent categorization \typo{in terms of} the spatial extent and characterization of the water masses (Figure \ref{figure_comparison1}). In particular, ALC-S can delineate two small regions (Gulf of Genoa and Strait of Gibraltar) with \typo{distinct} temperature patterns.}

\paolo{We further analyzed the classification results of several clustering methods that exhibited excellent performance in the previous simulations section (Section \ref{sec4}). We selected the hierarchical clustering method based on raw data, the GSC algorithm, and the FPCAC procedure using $93$ B-spline \typo{bases} (3 bases for each year). Although it is not a spatial method, it obtained high ARI values in the simulations.
Since these non-model-based methods were used, we selected the number of clusters that maximized the average silhouette width by utilizing the \texttt{fviz\_nbclust} function, which is \typo{available} in the R package \texttt{factoextra}. Please refer to Figure \ref{figure_waic} in the Supplementary Materials for additional information.}

\begin{figure}[ht!]
  \centering
\begin{tabular}{c}
\includegraphics[width=0.95\linewidth]{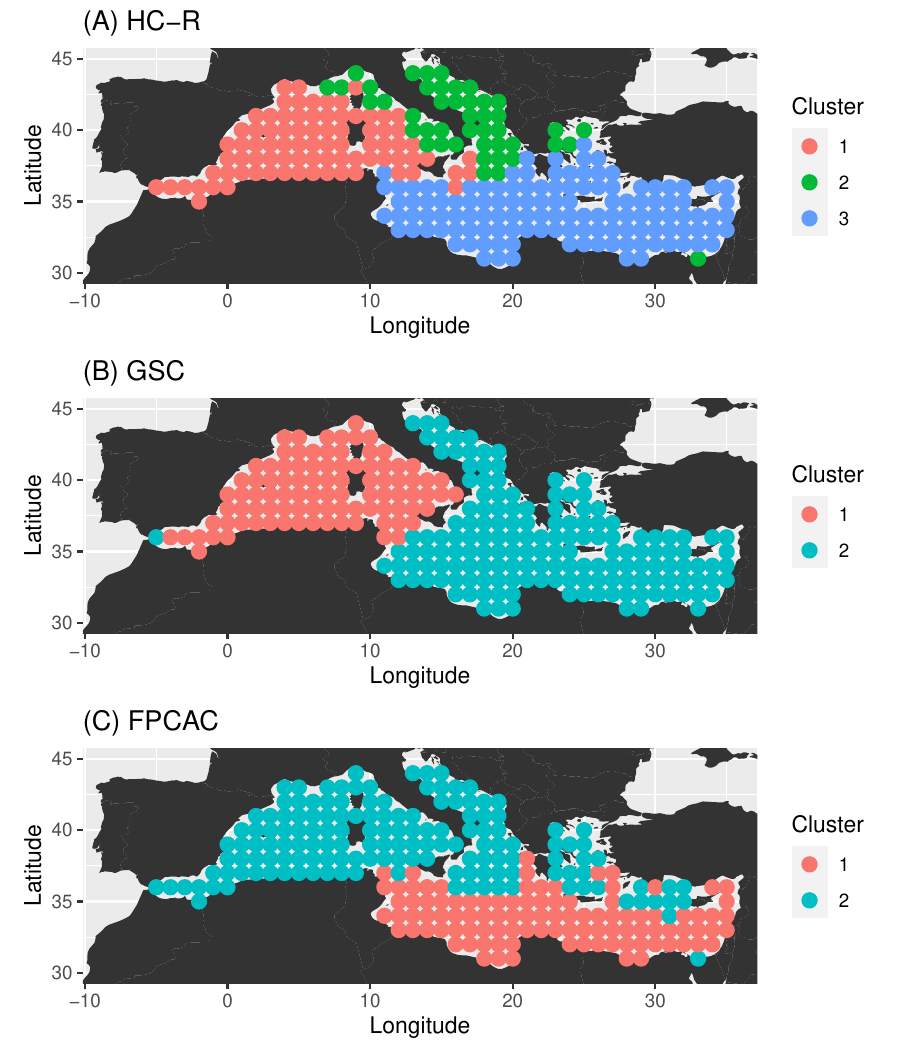}
\end{tabular}
	\caption{Spatial clustering by the three alternative clustering approaches\label{figure_comparison2}. }
\end{figure}

\paolo{Figure \ref{figure_comparison2} shows the spatial pattern of the three clustering methods considered. The partitions obtained by \typo{the} HC-R procedures suggest 3 clusters and largely separate the Mediterranean basin into three parts, resulting in a coarse classification (Figure 1A). The results obtained \typo{by} GSC and FPCAC suggest dividing the Mediterranean Sea into two groups based on their approaches: eastern vs. western for GSC and southern vs. northern for FPCAC. However, the GSC results \typo{appear} to be more spatially consistent.}

\begin{table}[ht]
\caption{\label{table_clustering}Summary statistics of the SST (in $\,^o$C) for each of the clusters identified by ALC-S procedure based on 8-NN network.}
\begin{tabular}{ccccccc}
\toprule
\textbf{Cluster} & \textbf{1} & \textbf{2} & \textbf{3} & \textbf{4} & \textbf{5} &\textbf{6}\\
\midrule
Size & 2 & 21 & 107 & 2 &24 &95\\ 
\hline
Average & 18.2 & 18.3 & 19.0 & 17.9 & 20.2 & 20.8 \\ 
Standard Deviation & 4.2 & 4.7 & 4.2 & 2.4 & 4.0 & 4.0 \\ 
Median & 17.7 & 17.7 & 18.3 & 17.4 & 19.7 & 20.4 \\ 
Median Abs. Deviation & 5.6 & 6.0 & 5.4 & 2.7 & 4.9 & 5.4 \\ 
Min & 12.4 & 8.2 & 11.4 & 14.2 & 12.5 & 12.9 \\ 
Max & 27.5 & 30.8 & 28.8 & 25.6 & 29.9 & 29.8 \\ 
Average 10-year increase °C& 0.32 &0.53&0.56& 0.43&0.48 &0.61\\
Median  intra-annual $\operatorname{IQR}$ &3.76 & 8.69 & 7.64 & 7.79 & 7.41 & 7.00\\
\bottomrule
\end{tabular}
\end{table}

\begin{figure}[ht]
  \centering
  \includegraphics[width=0.95\linewidth]{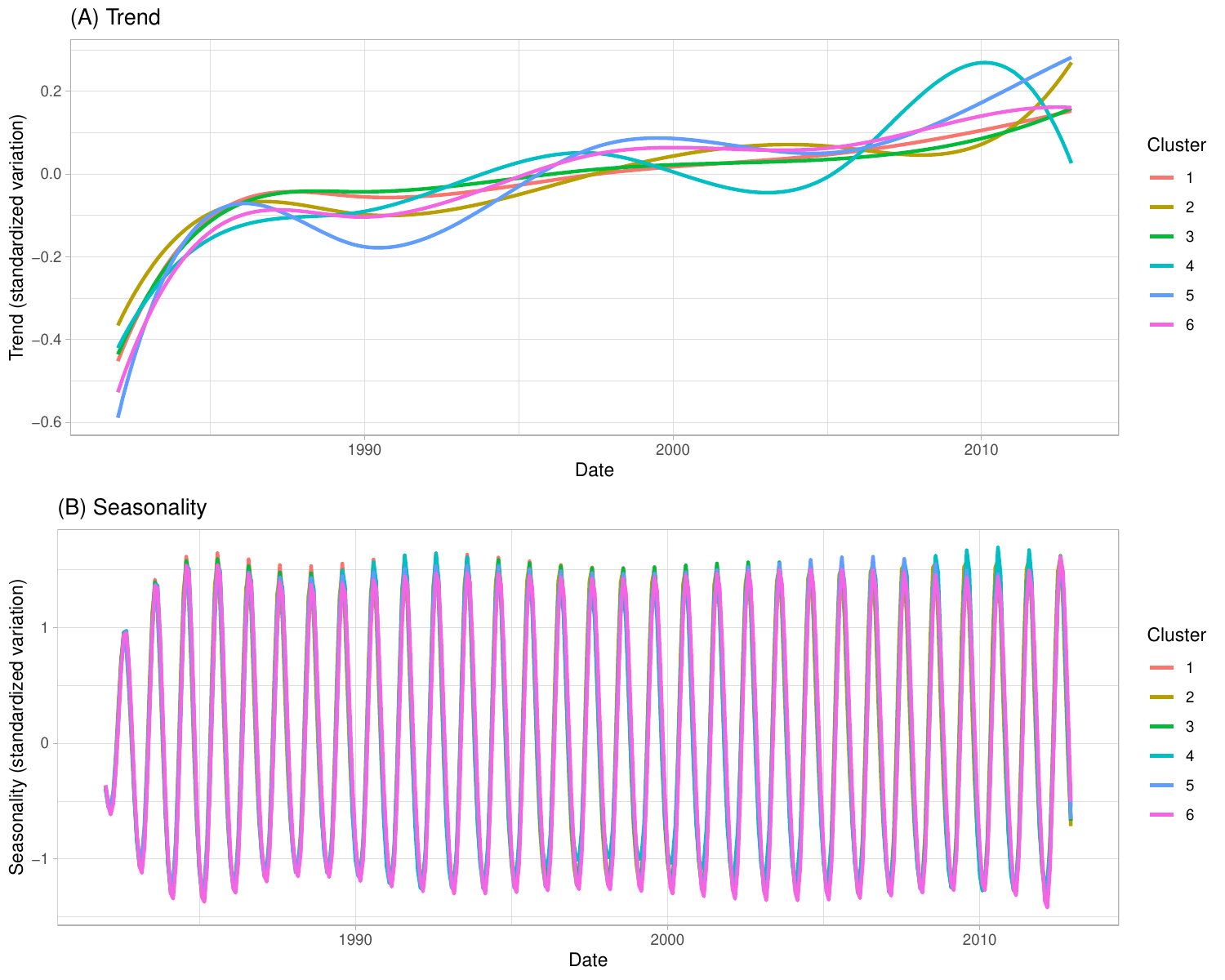}
	\caption{Estimated median values for the standardized temperature anomalies: trend (A) and seasonal (B) components for each cluster resulted by ALC-S procedure with 8-NN.\label{figure_curves} }
\end{figure}

To better understand the differences \typo{between} the defined groups of time series, \typo{the} position and variability indices, including the median intra-annual IQR, are reported in Table \ref{table_clustering}. For instance, Cluster $1$ and $4$ have a similar average SST value, but a marked difference in terms of range and intra-annual variation. Clusters $5$ and $6$ are spatially close in terms of average temperature, but \typo{show} different \typo{behavior} in terms of average temperature increase (0.48 vs. 0.61 °C decade$^{-1}$). The clusters $2$ and $3$ are very similar in terms of median/mean, but report a different seasonal pattern.
It is important to note that our classification is similar to that obtained by other studies considering a wide range of physical indicators \citep{delaHoz:2018} or looking at the trophic status based on chlorophyll type-a concentration \citep{d2008trophic}, showing how changes in temperature are related to many other environmental indicators. 

\begin{figure}[ht]
  \centering
  \includegraphics[width=0.95\linewidth]{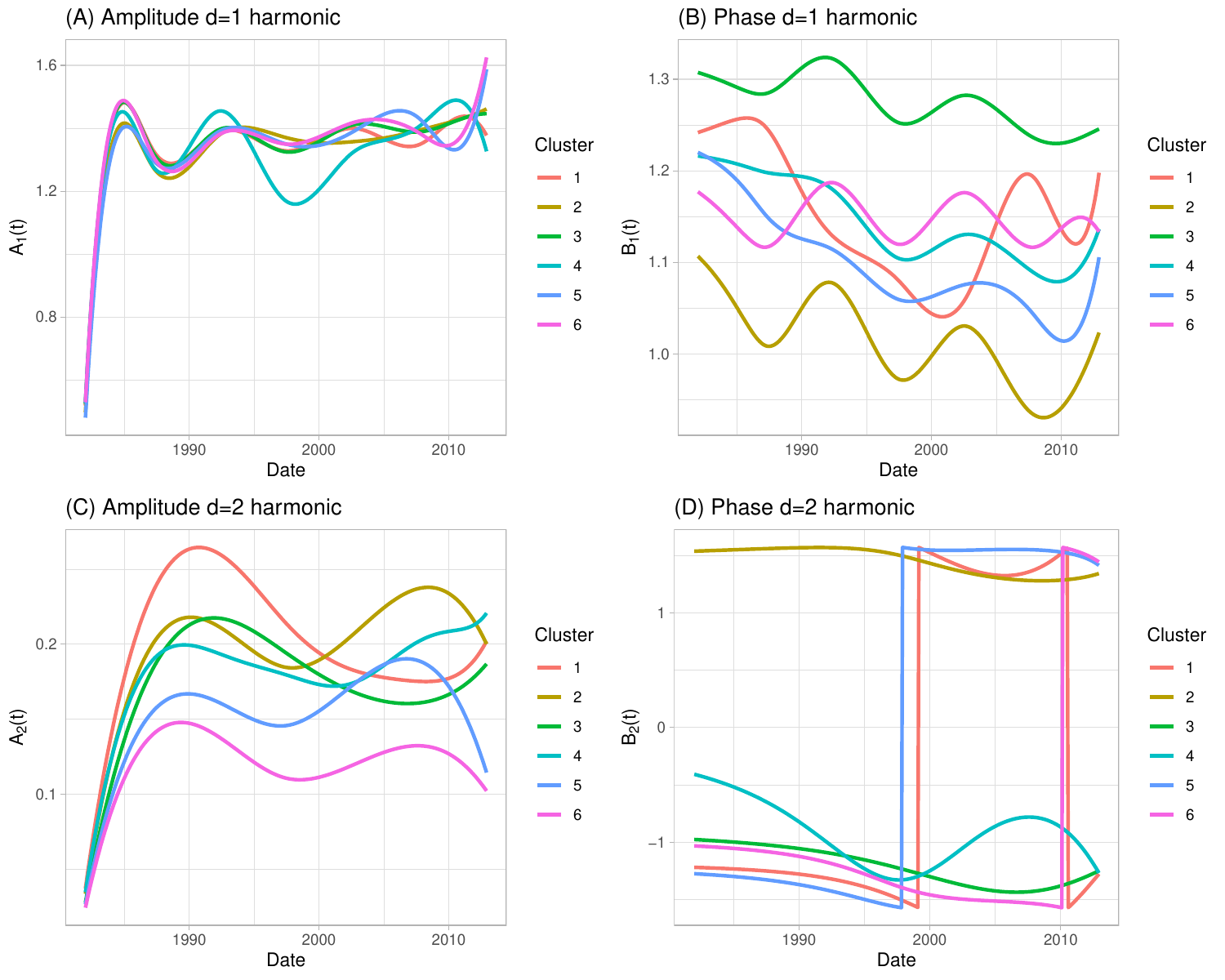}
	\caption{Amplitude and phase function of each harmonic estimated by ALC-S procedure with 8-NN.\label{figure_seas_component} }
\end{figure}

Looking at the median trend and seasonality curves shown in Figure \ref{figure_curves}, the overall temporal behavior is similar for all the identified clusters, with a steadily increasing trend with an overall increase close to \paolo{1 standard deviation. The most obvious difference concerns \typo{the} seasonality, but it seems to be too difficult to distinguish the main differences from Figure \ref{figure_curves}.  
By analyzing the seasonality decomposition in the first and second harmonics, we can extract the amplitude and phase variation through the following transformation: 
$$
A_d(t)=\sqrt{(\hat{g}_{2d}(t)^2+\hat{g}_{3d}(t)^2)},\quad B_d(t)=\arctan{(\hat{g}_{3d}(t)/\hat{g}_{2d}(t))}
$$
where $A_d(t)$ and $B_d(t)$ represent the functions related to the amplitude and phase, respectively, of the $d$-th harmonic. 
From the data in Figure \ref{figure_seas_component}, it is clear that the primary difference between the clusters is based on the phase functions that integrate \typo{the shift information of the seasonal component and the amplitude of the second harmonic}.}

\section{Discussion}\label{sec6}
In this paper, we propose a model for clustering time series based on quantiles. The model exploits a link between quantile regression and the likelihood of an asymmetric Laplace distribution. 
The model can capture the characteristics of data dispersion and detect the spatial extent of clusters, taking into account different spatial membership structures. 
In this paper, spatial dependence was incorporated using a Potts model of neighborhood membership. 
A Bayesian framework for deriving a posteriori estimates of the parameters is facilitated by the hierarchical structure.
In this regard, it is not difficult to extend the formulation to allow time series, within the same cluster, to be spatially and temporally dependent by assuming space-time correlated Gaussian errors in Equation \eqref{eq3}.
\paolo{Furthermore, our model formulation is not limited to temporal data but can be effortlessly expanded to spatial longitudinal data with covariates.}

\paolo{The analysis results of the Mediterranean Sea Surface Temperature are consistent with the literature, which classified sub-basins \citep{pastor:2019} or used abiotic characteristics and spatial hierarchical clustering of raw data \citep{delaHoz:2018}. Nevertheless, our clustering method provides a more coherent spatio-temporal classification of the water masses when compared to prior classification attempts. Our method \typo{allows} the characterization of both well-defined basins, such as the Adriatic Sea, and clusters with small spatial \typo{extent} and specific temporal \typo{behavior}, as indicated by the presence of fresh Atlantic waters in the Strait of Gibraltar \citep{bouzinac:2003} or the unique conformation of the Gulf of Genoa \citep{cutroneo2023cold}. Moreover, the classification presented here may \typo{help to identify} of water masses most affected by the \typo{effects} of climate change \citep{shaltout:2014}.
Although our clustering results are based on the median to estimate the central tendency of \typo{the} time series, an important advantage of our approach is the \typo{possibility} to obtain \typo{different} classifications by \typo{varying} the quantile of interest \typo{according to the objectives of the study}\citep{cade:2003}.}

Although the model has been adapted to cluster time series, the semi-parametric formulation makes it suitable for application to functional data as well. In this respect, a work closely related to ours on functional data clustering is \citet{kim-oh:2020}. However, the strength of our proposal is that it accounts for all \typo{uncertainties} in the clustering procedure. \paolo{An additional feature included is a flexible spatial dependence layer that \typo{allows} coherent classifications and the grading of spatial dependence strength within each cluster.}

The presented method has some limitations, mainly represented by the time required for the MCMC iterations. In fact, in our motivating example, which consists of 251 time series with 372 time points, it takes about \paolo{1,500} seconds to run 3,000 iterations with $K = 2$ clusters and a \paolo{31}-column regressor matrix. Using an ad hoc expectation maximization algorithm can be time-saving but with a reduced number of features that can be included in the model specification compared to our Bayesian framework. 
\paolo{Another interesting aspect pertains to the application of real data where the spatial domain is non-convex and characterized by a water distance that may not be Euclidean due to the presence of coasts and land between two locations. However, \typo{the specification of anisotropic models can be challenging and context dependent}. By implementing an inter-cluster, variable spatial dependence layer, one can achieve general adaptability in numerous real-data applications, despite the use of a standard isotropic Potts model.}
Another limitation is that only one quantile can be considered. Extending this work to a context similar to \citet{bondell-reich-huixia:2010}\typo{'s} is a research direction to be explored.

\section*{Acknowledgement} The authors would like to thank No\'emie Le Carrer for retrieving the dataset.
\newpage
\section*{Declarations}
\subsection{Funding}
This research received no specific grant from any funding agency in the public, commercial, or not-for-profit sectors.

\subsection{Conflict of interest}
The authors have no conflicts of interest to declare.

\subsection{Code and data availability}
Code and data are available upon request from the authors

\newpage
\bibliography{myref-4}


\begin{thebibliography}{67}
\ifx \bisbn   \undefined \def \bisbn  #1{ISBN #1}\fi
\ifx \binits  \undefined \def \binits#1{#1}\fi
\ifx \bauthor  \undefined \def \bauthor#1{#1}\fi
\ifx \batitle  \undefined \def \batitle#1{#1}\fi
\ifx \bjtitle  \undefined \def \bjtitle#1{#1}\fi
\ifx \bvolume  \undefined \def \bvolume#1{\textbf{#1}}\fi
\ifx \byear  \undefined \def \byear#1{#1}\fi
\ifx \bissue  \undefined \def \bissue#1{#1}\fi
\ifx \bfpage  \undefined \def \bfpage#1{#1}\fi
\ifx \blpage  \undefined \def \blpage #1{#1}\fi
\ifx \burl  \undefined \def \burl#1{\textsf{#1}}\fi
\ifx \doiurl  \undefined \def \doiurl#1{\url{https://doi.org/#1}}\fi
\ifx \betal  \undefined \def \betal{\textit{et al.}}\fi
\ifx \binstitute  \undefined \def \binstitute#1{#1}\fi
\ifx \binstitutionaled  \undefined \def \binstitutionaled#1{#1}\fi
\ifx \bctitle  \undefined \def \bctitle#1{#1}\fi
\ifx \beditor  \undefined \def \beditor#1{#1}\fi
\ifx \bpublisher  \undefined \def \bpublisher#1{#1}\fi
\ifx \bbtitle  \undefined \def \bbtitle#1{#1}\fi
\ifx \bedition  \undefined \def \bedition#1{#1}\fi
\ifx \bseriesno  \undefined \def \bseriesno#1{#1}\fi
\ifx \blocation  \undefined \def \blocation#1{#1}\fi
\ifx \bsertitle  \undefined \def \bsertitle#1{#1}\fi
\ifx \bsnm \undefined \def \bsnm#1{#1}\fi
\ifx \bsuffix \undefined \def \bsuffix#1{#1}\fi
\ifx \bparticle \undefined \def \bparticle#1{#1}\fi
\ifx \barticle \undefined \def \barticle#1{#1}\fi
\bibcommenthead
\ifx \bconfdate \undefined \def \bconfdate #1{#1}\fi
\ifx \botherref \undefined \def \botherref #1{#1}\fi
\ifx \url \undefined \def \url#1{\textsf{#1}}\fi
\ifx \bchapter \undefined \def \bchapter#1{#1}\fi
\ifx \bbook \undefined \def \bbook#1{#1}\fi
\ifx \bcomment \undefined \def \bcomment#1{#1}\fi
\ifx \oauthor \undefined \def \oauthor#1{#1}\fi
\ifx \citeauthoryear \undefined \def \citeauthoryear#1{#1}\fi
\ifx \endbibitem  \undefined \def \endbibitem {}\fi
\ifx \bconflocation  \undefined \def \bconflocation#1{#1}\fi
\ifx \arxivurl  \undefined \def \arxivurl#1{\textsf{#1}}\fi
\csname PreBibitemsHook\endcsname

\bibitem[\protect\citeauthoryear{Schneider}{2001}]{Schneider:2001}
\begin{barticle}
\bauthor{\bsnm{Schneider}, \binits{S.H.}}:
\batitle{What is 'dangerous' climate change?}
\bjtitle{Nature}
\bvolume{411},
\bfpage{17}--\blpage{19}
(\byear{2001})
\end{barticle}
\endbibitem

\bibitem[\protect\citeauthoryear{Portmann et~al.}{2009}]{Portmann:2009}
\begin{barticle}
\bauthor{\bsnm{Portmann}, \binits{R.W.}},
\bauthor{\bsnm{Solomon}, \binits{S.}},
\bauthor{\bsnm{Hegerl}, \binits{G.C.}}:
\batitle{Spatial and seasonal patterns in climate change, temperatures, and
  precipitation across the {U}nited {S}tates}.
\bjtitle{Proceedings of the National Academy of Sciences}
\bvolume{106},
\bfpage{7324}--\blpage{7329}
(\byear{2009})
\end{barticle}
\endbibitem

\bibitem[\protect\citeauthoryear{Katz}{2010}]{Katz:2010}
\begin{barticle}
\bauthor{\bsnm{Katz}, \binits{R.W.}}:
\batitle{Statistics of extremes in climate change}.
\bjtitle{Climatic Change}
\bvolume{100},
\bfpage{71}--\blpage{76}
(\byear{2010})
\end{barticle}
\endbibitem

\bibitem[\protect\citeauthoryear{Sun et~al.}{2018}]{Sun:2018}
\begin{barticle}
\bauthor{\bsnm{Sun}, \binits{F.}},
\bauthor{\bsnm{Roderick}, \binits{M.L.}},
\bauthor{\bsnm{Farquhar}, \binits{G.D.}}:
\batitle{Rainfall statistics, stationarity, and climate change}.
\bjtitle{Proceedings of the National Academy of Sciences}
\bvolume{115},
\bfpage{2305}--\blpage{2310}
(\byear{2018})
\end{barticle}
\endbibitem

\bibitem[\protect\citeauthoryear{Reich}{2012}]{reich:2012}
\begin{barticle}
\bauthor{\bsnm{Reich}, \binits{B.J.}}:
\batitle{Spatiotemporal quantile regression for detecting distributional
  changes in environmental processes}.
\bjtitle{Journal of the Royal Statistical Society Series C: Applied Statistics}
\bvolume{61},
\bfpage{535}--\blpage{553}
(\byear{2012})
\end{barticle}
\endbibitem

\bibitem[\protect\citeauthoryear{Cannon}{2018}]{cannon:2018}
\begin{barticle}
\bauthor{\bsnm{Cannon}, \binits{A.J.}}:
\batitle{Non-crossing nonlinear regression quantiles by monotone composite
  quantile regression neural network, with application to rainfall extremes}.
\bjtitle{Stochastic Environmental Research and Risk Assessment}
\bvolume{32},
\bfpage{3207}--\blpage{3225}
(\byear{2018})
\end{barticle}
\endbibitem

\bibitem[\protect\citeauthoryear{Vandeskog et~al.}{2022}]{Vandeskog-et-al:2022}
\begin{barticle}
\bauthor{\bsnm{Vandeskog}, \binits{S.M.}},
\bauthor{\bsnm{Thorarinsdottir}, \binits{T.L.}},
\bauthor{\bsnm{Steinsland}, \binits{I.}},
\bauthor{\bsnm{Lindgren}, \binits{F.}}:
\batitle{Quantile based modeling of diurnal temperature range with the
  five-parameter lambda distribution}.
\bjtitle{Environmetrics}
\bvolume{33},
\bfpage{2719}
(\byear{2022})
\end{barticle}
\endbibitem

\bibitem[\protect\citeauthoryear{Liao}{2005}]{Liao2005}
\begin{barticle}
\bauthor{\bsnm{Liao}, \binits{T.W.}}:
\batitle{Clustering of time series data—a survey}.
\bjtitle{Pattern Recognition}
\bvolume{38},
\bfpage{1857}--\blpage{1874}
(\byear{2005})
\end{barticle}
\endbibitem

\bibitem[\protect\citeauthoryear{Zhang and Parnell}{2023}]{Zhang:Parnell:2023}
\begin{botherref}
\oauthor{\bsnm{Zhang}, \binits{M.}},
\oauthor{\bsnm{Parnell}, \binits{A.}}:
Review of clustering methods for functional data.
ACM Transactions on Knowledge Discovery from Data
\textbf{17}
(2023)
\end{botherref}
\endbibitem

\bibitem[\protect\citeauthoryear{Gr{\"u}n and Leisch}{2007}]{grun2007fitting}
\begin{barticle}
\bauthor{\bsnm{Gr{\"u}n}, \binits{B.}},
\bauthor{\bsnm{Leisch}, \binits{F.}}:
\batitle{Fitting finite mixtures of generalized linear regressions in r}.
\bjtitle{Computational Statistics \& Data Analysis}
\bvolume{51},
\bfpage{5247}--\blpage{5252}
(\byear{2007})
\end{barticle}
\endbibitem

\bibitem[\protect\citeauthoryear{Jiang and Serban}{2012}]{Jiang2012}
\begin{barticle}
\bauthor{\bsnm{Jiang}, \binits{H.}},
\bauthor{\bsnm{Serban}, \binits{N.}}:
\batitle{Clustering random curves under spatial interdependence with
  application to service accessibility}.
\bjtitle{Technometrics}
\bvolume{54},
\bfpage{108}--\blpage{119}
(\byear{2012})
\end{barticle}
\endbibitem

\bibitem[\protect\citeauthoryear{Cuesta-Albertos
  et~al.}{1997}]{cuesta1997trimmed}
\begin{barticle}
\bauthor{\bsnm{Cuesta-Albertos}, \binits{J.A.}},
\bauthor{\bsnm{Gordaliza}, \binits{A.}},
\bauthor{\bsnm{Matr{\'a}n}, \binits{C.}}:
\batitle{Trimmed $ k $-means: an attempt to robustify quantizers}.
\bjtitle{The Annals of Statistics}
\bvolume{25},
\bfpage{553}--\blpage{576}
(\byear{1997})
\end{barticle}
\endbibitem

\bibitem[\protect\citeauthoryear{Gallegos}{2002}]{gallegos2002maximum}
\begin{bchapter}
\bauthor{\bsnm{Gallegos}, \binits{M.T.}}:
\bctitle{Maximum likelihood clustering with outliers}.
In: \beditor{\bsnm{Jajuga}, \binits{K.}},
\beditor{\bsnm{Soko{\l}owski}, \binits{A.}},
\beditor{\bsnm{Bock}, \binits{H.-H.}} (eds.)
\bbtitle{Classification, Clustering, and Data Analysis},
pp. \bfpage{247}--\blpage{255}.
\bpublisher{Springer},
\blocation{Berlin, Heidelberg}
(\byear{2002})
\end{bchapter}
\endbibitem

\bibitem[\protect\citeauthoryear{Amovin-Assagba et~al.}{2022}]{amovin:2022}
\begin{barticle}
\bauthor{\bsnm{Amovin-Assagba}, \binits{M.}},
\bauthor{\bsnm{Gannaz}, \binits{I.}},
\bauthor{\bsnm{Jacques}, \binits{J.}}:
\batitle{Outlier detection in multivariate functional data through a
  contaminated mixture model}.
\bjtitle{Computational Statistics \& Data Analysis}
\bvolume{174},
\bfpage{107496}
(\byear{2022})
\end{barticle}
\endbibitem

\bibitem[\protect\citeauthoryear{Garcia-Escudero
  et~al.}{2015}]{garcia2015avoiding}
\begin{barticle}
\bauthor{\bsnm{Garcia-Escudero}, \binits{L.A.}},
\bauthor{\bsnm{Gordaliza}, \binits{A.}},
\bauthor{\bsnm{Matr{\'a}n}, \binits{C.}},
\bauthor{\bsnm{Mayo-Iscar}, \binits{A.}}:
\batitle{Avoiding spurious local maximizers in mixture modeling}.
\bjtitle{Statistics and Computing}
\bvolume{25},
\bfpage{619}--\blpage{633}
(\byear{2015})
\end{barticle}
\endbibitem

\bibitem[\protect\citeauthoryear{Fritz et~al.}{2013}]{fritz2013fast}
\begin{barticle}
\bauthor{\bsnm{Fritz}, \binits{H.}},
\bauthor{\bsnm{Garcia-Escudero}, \binits{L.A.}},
\bauthor{\bsnm{Mayo-Iscar}, \binits{A.}}:
\batitle{A fast algorithm for robust constrained clustering}.
\bjtitle{Computational Statistics \& Data Analysis}
\bvolume{61},
\bfpage{124}--\blpage{136}
(\byear{2013})
\end{barticle}
\endbibitem

\bibitem[\protect\citeauthoryear{Nguyen et~al.}{2016}]{nguyen2016spatial}
\begin{barticle}
\bauthor{\bsnm{Nguyen}, \binits{H.D.}},
\bauthor{\bsnm{McLachlan}, \binits{G.J.}},
\bauthor{\bsnm{Ullmann}, \binits{J.F.}},
\bauthor{\bsnm{Janke}, \binits{A.L.}}:
\batitle{Spatial clustering of time series via mixture of autoregressions
  models and {M}arkov random fields}.
\bjtitle{Statistica Neerlandica}
\bvolume{70},
\bfpage{414}--\blpage{439}
(\byear{2016})
\end{barticle}
\endbibitem

\bibitem[\protect\citeauthoryear{Disegna et~al.}{2017}]{disegna2017copula}
\begin{barticle}
\bauthor{\bsnm{Disegna}, \binits{M.}},
\bauthor{\bsnm{D’Urso}, \binits{P.}},
\bauthor{\bsnm{Durante}, \binits{F.}}:
\batitle{Copula-based fuzzy clustering of spatial time series}.
\bjtitle{Spatial Statistics}
\bvolume{21},
\bfpage{209}--\blpage{225}
(\byear{2017})
\end{barticle}
\endbibitem

\bibitem[\protect\citeauthoryear{Delicado
  et~al.}{2010}]{delicado2010statistics}
\begin{barticle}
\bauthor{\bsnm{Delicado}, \binits{P.}},
\bauthor{\bsnm{Giraldo}, \binits{R.}},
\bauthor{\bsnm{Comas}, \binits{C.}},
\bauthor{\bsnm{Mateu}, \binits{J.}}:
\batitle{Statistics for spatial functional data: some recent contributions}.
\bjtitle{Environmetrics}
\bvolume{21},
\bfpage{224}--\blpage{239}
(\byear{2010})
\end{barticle}
\endbibitem

\bibitem[\protect\citeauthoryear{Vandewalle
  et~al.}{2022}]{vandewalle2022clustering}
\begin{bchapter}
\bauthor{\bsnm{Vandewalle}, \binits{V.}},
\bauthor{\bsnm{Preda}, \binits{C.}},
\bauthor{\bsnm{Dabo-Niang}, \binits{S.}}:
\bctitle{Clustering spatial functional data}.
In: \beditor{\bsnm{Mateu}, \binits{J.}},
\beditor{\bsnm{Giraldo}, \binits{R.}} (eds.)
\bbtitle{Geostatistical Functional Data Analysis},
pp. \bfpage{155}--\blpage{174}.
\bpublisher{Wiley},
\blocation{New York}
(\byear{2022})
\end{bchapter}
\endbibitem

\bibitem[\protect\citeauthoryear{Koner and Staicu}{2023}]{koner:2023}
\begin{barticle}
\bauthor{\bsnm{Koner}, \binits{S.}},
\bauthor{\bsnm{Staicu}, \binits{A.-M.}}:
\batitle{Second-generation functional data}.
\bjtitle{Annual Review of Statistics and Its Application}
\bvolume{10},
\bfpage{547}--\blpage{572}
(\byear{2023})
\end{barticle}
\endbibitem

\bibitem[\protect\citeauthoryear{Jiang and Serban}{2012}]{jiang2012clustering}
\begin{barticle}
\bauthor{\bsnm{Jiang}, \binits{H.}},
\bauthor{\bsnm{Serban}, \binits{N.}}:
\batitle{Clustering random curves under spatial interdependence with
  application to service accessibility}.
\bjtitle{Technometrics}
\bvolume{54},
\bfpage{108}--\blpage{119}
(\byear{2012})
\end{barticle}
\endbibitem

\bibitem[\protect\citeauthoryear{Hu et~al.}{2022}]{hu2022bayesian}
\begin{barticle}
\bauthor{\bsnm{Hu}, \binits{G.}},
\bauthor{\bsnm{Geng}, \binits{J.}},
\bauthor{\bsnm{Xue}, \binits{Y.}},
\bauthor{\bsnm{Sang}, \binits{H.}}:
\batitle{Bayesian spatial homogeneity pursuit of functional data: an
  application to the us income distribution}.
\bjtitle{Bayesian Analysis}
\bvolume{1},
\bfpage{1}--\blpage{27}
(\byear{2022})
\end{barticle}
\endbibitem

\bibitem[\protect\citeauthoryear{Secchi
  et~al.}{2013}]{Secchi:Vantini:Vitelli:2013}
\begin{barticle}
\bauthor{\bsnm{Secchi}, \binits{P.}},
\bauthor{\bsnm{Vantini}, \binits{S.}},
\bauthor{\bsnm{Vitelli}, \binits{V.}}:
\batitle{Bagging {V}oronoi classifiers for clustering spatial functional data}.
\bjtitle{International Journal of Applied Earth Observation and Geoinformation}
\bvolume{22},
\bfpage{53}--\blpage{64}
(\byear{2013})
\end{barticle}
\endbibitem

\bibitem[\protect\citeauthoryear{Giraldo
  et~al.}{2012}]{Giraldo:Delicado:Mateu:2012}
\begin{barticle}
\bauthor{\bsnm{Giraldo}, \binits{R.}},
\bauthor{\bsnm{Delicado}, \binits{P.}},
\bauthor{\bsnm{Mateu}, \binits{J.}}:
\batitle{Hierarchical clustering of spatially correlated functional data}.
\bjtitle{Statistica Neerlandica}
\bvolume{66},
\bfpage{403}--\blpage{421}
(\byear{2012})
\end{barticle}
\endbibitem

\bibitem[\protect\citeauthoryear{Bera et~al.}{2016}]{bera:2016}
\begin{barticle}
\bauthor{\bsnm{Bera}, \binits{A.K.}},
\bauthor{\bsnm{Galvao~Jr}, \binits{A.F.}},
\bauthor{\bsnm{Montes-Rojas}, \binits{G.V.}},
\bauthor{\bsnm{Park}, \binits{S.Y.}}:
\batitle{Asymmetric {L}aplace regression: maximum likelihood, maximum entropy
  and quantile regression}.
\bjtitle{Journal of Econometric Methods}
\bvolume{5},
\bfpage{79}--\blpage{101}
(\byear{2016})
\end{barticle}
\endbibitem

\bibitem[\protect\citeauthoryear{Potts}{1952}]{Potts:1952}
\begin{barticle}
\bauthor{\bsnm{Potts}, \binits{R.B.}}:
\batitle{Some generalized order-disorder transformations}.
\bjtitle{Mathematical Proceedings of the Cambridge Philosophical Society}
\bvolume{48},
\bfpage{106}--\blpage{109}
(\byear{1952})
\end{barticle}
\endbibitem

\bibitem[\protect\citeauthoryear{Nykjaer}{2009}]{nykjaer:2009}
\begin{barticle}
\bauthor{\bsnm{Nykjaer}, \binits{L.}}:
\batitle{Mediterranean {S}ea surface warming 1985--2006}.
\bjtitle{Climate Research}
\bvolume{39},
\bfpage{11}--\blpage{17}
(\byear{2009})
\end{barticle}
\endbibitem

\bibitem[\protect\citeauthoryear{Lejeusne et~al.}{2010}]{lejeusne:2010}
\begin{barticle}
\bauthor{\bsnm{Lejeusne}, \binits{C.}},
\bauthor{\bsnm{Chevaldonn{\'e}}, \binits{P.}},
\bauthor{\bsnm{Pergent-Martini}, \binits{C.}},
\bauthor{\bsnm{Boudouresque}, \binits{C.F.}},
\bauthor{\bsnm{P{\'e}rez}, \binits{T.}}:
\batitle{Climate change effects on a miniature ocean: the highly diverse,
  highly impacted {M}editerranean {S}ea}.
\bjtitle{Trends in Ecology \& Evolution}
\bvolume{25},
\bfpage{250}--\blpage{260}
(\byear{2010})
\end{barticle}
\endbibitem

\bibitem[\protect\citeauthoryear{Bethoux et~al.}{1990}]{bethoux:1990}
\begin{barticle}
\bauthor{\bsnm{Bethoux}, \binits{J.-P.}},
\bauthor{\bsnm{Gentili}, \binits{B.}},
\bauthor{\bsnm{Raunet}, \binits{J.}},
\bauthor{\bsnm{Tailliez}, \binits{D.}}:
\batitle{Warming trend in the western {M}editerranean deep water}.
\bjtitle{Nature}
\bvolume{347},
\bfpage{660}--\blpage{662}
(\byear{1990})
\end{barticle}
\endbibitem

\bibitem[\protect\citeauthoryear{Ibrahim et~al.}{2021}]{ibrahim:2021}
\begin{barticle}
\bauthor{\bsnm{Ibrahim}, \binits{O.}},
\bauthor{\bsnm{Mohamed}, \binits{B.}},
\bauthor{\bsnm{Nagy}, \binits{H.}}:
\batitle{Spatial variability and trends of marine heat waves in the {e}astern
  {M}editerranean {S}ea over 39 years}.
\bjtitle{Journal of Marine Science and Engineering}
\bvolume{9},
\bfpage{643}
(\byear{2021})
\end{barticle}
\endbibitem

\bibitem[\protect\citeauthoryear{Nunes et~al.}{2019}]{nunes:2019}
\begin{barticle}
\bauthor{\bsnm{Nunes}, \binits{S.}},
\bauthor{\bsnm{Perez}, \binits{G.L.}},
\bauthor{\bsnm{Latasa}, \binits{M.}},
\bauthor{\bsnm{Zamanillo}, \binits{M.}},
\bauthor{\bsnm{Delgado}, \binits{M.}},
\bauthor{\bsnm{Ortega-Retuerta}, \binits{E.}},
\bauthor{\bsnm{Marras\'e}, \binits{C.}},
\bauthor{\bsnm{Sim\'o}, \binits{R.}},
\bauthor{\bsnm{Estrada}, \binits{M.}}:
\batitle{Size fractionation, chemotaxonomic groups and bio-optical properties
  of phytoplankton along a transect from the {Mediterranean Sea} to the {SW
  Atlantic Ocean}}.
\bjtitle{Scientia Marina}
\bvolume{83},
\bfpage{87}--\blpage{109}
(\byear{2019})
\end{barticle}
\endbibitem

\bibitem[\protect\citeauthoryear{Pastor et~al.}{2019}]{pastor:2019}
\begin{barticle}
\bauthor{\bsnm{Pastor}, \binits{F.}},
\bauthor{\bsnm{Valiente}, \binits{J.A.}},
\bauthor{\bsnm{Palau}, \binits{J.L.}}:
\batitle{Sea surface temperature in the mediterranean: Trends and spatial
  patterns (1982--2016)}.
\bjtitle{Meteorology and Climatology of the Mediterranean and Black Seas}
\bvolume{175},
\bfpage{297}--\blpage{309}
(\byear{2019})
\end{barticle}
\endbibitem

\bibitem[\protect\citeauthoryear{McLachlan and
  Peel}{2000}]{mclachlan:peel:2000}
\begin{bbook}
\bauthor{\bsnm{McLachlan}, \binits{G.J.}},
\bauthor{\bsnm{Peel}, \binits{D.}}:
\bbtitle{Finite {M}ixture {M}odels}.
\bpublisher{Wiley},
\blocation{New York}
(\byear{2000})
\end{bbook}
\endbibitem

\bibitem[\protect\citeauthoryear{Strauss}{1977}]{Strauss:1977}
\begin{barticle}
\bauthor{\bsnm{Strauss}, \binits{D.J.}}:
\batitle{Clustering on coloured lattices}.
\bjtitle{Journal of Applied Probability}
\bvolume{14},
\bfpage{135}--\blpage{143}
(\byear{1977})
\end{barticle}
\endbibitem

\bibitem[\protect\citeauthoryear{Gaetan et~al.}{2017}]{Gaetan:2017}
\begin{barticle}
\bauthor{\bsnm{Gaetan}, \binits{C.}},
\bauthor{\bsnm{Girardi}, \binits{P.}},
\bauthor{\bsnm{Pastres}, \binits{R.}}:
\batitle{Spatial clustering of curves with an application of satellite data}.
\bjtitle{Spatial Statistics}
\bvolume{20},
\bfpage{110}--\blpage{124}
(\byear{2017})
\end{barticle}
\endbibitem

\bibitem[\protect\citeauthoryear{Koenker and
  Bassett}{1978}]{koenker1978regression}
\begin{barticle}
\bauthor{\bsnm{Koenker}, \binits{R.}},
\bauthor{\bsnm{Bassett}, \binits{G.}}:
\batitle{Regression quantiles}.
\bjtitle{Econometrica}
\bvolume{46},
\bfpage{33}--\blpage{50}
(\byear{1978})
\end{barticle}
\endbibitem

\bibitem[\protect\citeauthoryear{Koenker and
  Machado}{1999}]{koenker1999goodness}
\begin{barticle}
\bauthor{\bsnm{Koenker}, \binits{R.}},
\bauthor{\bsnm{Machado}, \binits{J.A.}}:
\batitle{Goodness of fit and related inference processes for quantile
  regression}.
\bjtitle{Journal of the American Statistical Association}
\bvolume{94},
\bfpage{1296}--\blpage{1310}
(\byear{1999})
\end{barticle}
\endbibitem

\bibitem[\protect\citeauthoryear{Huber}{1981}]{Huber:1981}
\begin{bbook}
\bauthor{\bsnm{Huber}, \binits{P.J.}}:
\bbtitle{Robust Statistics}.
\bpublisher{Wiley},
\blocation{New York}
(\byear{1981})
\end{bbook}
\endbibitem

\bibitem[\protect\citeauthoryear{Besag}{1986}]{Besag:1986}
\begin{barticle}
\bauthor{\bsnm{Besag}, \binits{J.}}:
\batitle{On the statistical analysis of dirty pictures}.
\bjtitle{Journal of the Royal Statistical Society: Series B (Methodological)}
\bvolume{48},
\bfpage{259}--\blpage{279}
(\byear{1986})
\end{barticle}
\endbibitem

\bibitem[\protect\citeauthoryear{Cucala et~al.}{2009}]{cucala_et_al:2009}
\begin{barticle}
\bauthor{\bsnm{Cucala}, \binits{L.}},
\bauthor{\bsnm{Marin}, \binits{J.-M.}},
\bauthor{\bsnm{Robert}, \binits{C.P.}},
\bauthor{\bsnm{Titterington}, \binits{D.M.}}:
\batitle{A {B}ayesian reassessment of nearest-neighbor classification}.
\bjtitle{Journal of the American Statistical Association}
\bvolume{104},
\bfpage{263}--\blpage{273}
(\byear{2009})
\end{barticle}
\endbibitem

\bibitem[\protect\citeauthoryear{Benoit and {Van den Poel}}{2017}]{Benoit2017}
\begin{barticle}
\bauthor{\bsnm{Benoit}, \binits{D.F.}},
\bauthor{\bsnm{{Van den Poel}}, \binits{D.}}:
\batitle{{bayesQR}: {A} {B}ayesian approach to quantile regression}.
\bjtitle{Journal of Statistical Software}
\bvolume{76},
\bfpage{1}--\blpage{32}
(\byear{2017})
\end{barticle}
\endbibitem

\bibitem[\protect\citeauthoryear{Robert and Casella}{2004}]{Robert2004}
\begin{bbook}
\bauthor{\bsnm{Robert}, \binits{C.P.}},
\bauthor{\bsnm{Casella}, \binits{G.}}:
\bbtitle{Monte Carlo Statistical Methods}.
\bpublisher{Springer},
\blocation{New York}
(\byear{2004})
\end{bbook}
\endbibitem

\bibitem[\protect\citeauthoryear{Kotz et~al.}{2001}]{Kotz2001}
\begin{bbook}
\bauthor{\bsnm{Kotz}, \binits{S.}},
\bauthor{\bsnm{Kozubowski}, \binits{T.}},
\bauthor{\bsnm{Podgorski}, \binits{K.}}:
\bbtitle{The {L}aplace Distribution and Generalizations: A Revisit with
  Applications to Communications, Economics, Engineering, and Finance}.
\bpublisher{Springer},
\blocation{New York}
(\byear{2001})
\end{bbook}
\endbibitem

\bibitem[\protect\citeauthoryear{Kozumi and Kobayashi}{2011}]{kozumi2011gibbs}
\begin{barticle}
\bauthor{\bsnm{Kozumi}, \binits{H.}},
\bauthor{\bsnm{Kobayashi}, \binits{G.}}:
\batitle{Gibbs sampling methods for {B}ayesian quantile regression}.
\bjtitle{Journal of Statistical Computation and Simulation}
\bvolume{81},
\bfpage{1565}--\blpage{1578}
(\byear{2011})
\end{barticle}
\endbibitem

\bibitem[\protect\citeauthoryear{Jorgensen}{1982}]{Jorgensen1982}
\begin{bbook}
\bauthor{\bsnm{Jorgensen}, \binits{B.}}:
\bbtitle{Statistical {P}roperties of the {G}eneralized {I}nverse {G}aussian
  {D}istribution}.
\bpublisher{Springer},
\blocation{New York}
(\byear{1982})
\end{bbook}
\endbibitem

\bibitem[\protect\citeauthoryear{Marjoram et~al.}{2003}]{marjoram2003markov}
\begin{barticle}
\bauthor{\bsnm{Marjoram}, \binits{P.}},
\bauthor{\bsnm{Molitor}, \binits{J.}},
\bauthor{\bsnm{Plagnol}, \binits{V.}},
\bauthor{\bsnm{Tavar{\'e}}, \binits{S.}}:
\batitle{Markov chain {M}onte {C}arlo without likelihoods}.
\bjtitle{Proceedings of the National Academy of Sciences}
\bvolume{100},
\bfpage{15324}--\blpage{15328}
(\byear{2003})
\end{barticle}
\endbibitem

\bibitem[\protect\citeauthoryear{Marin et~al.}{2012}]{marin2012approximate}
\begin{barticle}
\bauthor{\bsnm{Marin}, \binits{J.-M.}},
\bauthor{\bsnm{Pudlo}, \binits{P.}},
\bauthor{\bsnm{Robert}, \binits{C.P.}},
\bauthor{\bsnm{Ryder}, \binits{R.J.}}:
\batitle{Approximate {B}ayesian computational methods}.
\bjtitle{Statistics and Computing}
\bvolume{22},
\bfpage{1167}--\blpage{1180}
(\byear{2012})
\end{barticle}
\endbibitem

\bibitem[\protect\citeauthoryear{Pereyra et~al.}{2013}]{Pereyra:2013}
\begin{barticle}
\bauthor{\bsnm{Pereyra}, \binits{M.}},
\bauthor{\bsnm{Dobigeon}, \binits{N.}},
\bauthor{\bsnm{Batatia}, \binits{H.}},
\bauthor{\bsnm{Tourneret}, \binits{J.-Y.}}:
\batitle{Estimating the granularity coefficient of a {Potts-Markov} random
  field within a {Markov chain Monte Carlo} algorithm}.
\bjtitle{IEEE Transactions on Image Processing}
\bvolume{22},
\bfpage{2385}--\blpage{2397}
(\byear{2013})
\end{barticle}
\endbibitem

\bibitem[\protect\citeauthoryear{Sottile and Adelfio}{2019}]{Sottile:2019}
\begin{barticle}
\bauthor{\bsnm{Sottile}, \binits{G.}},
\bauthor{\bsnm{Adelfio}, \binits{G.}}:
\batitle{Clusters of effects curves in quantile regression models}.
\bjtitle{Computational Statistics}
\bvolume{34},
\bfpage{551}--\blpage{569}
(\byear{2019})
\end{barticle}
\endbibitem

\bibitem[\protect\citeauthoryear{Hubert and Arabie}{1985}]{Hubert1985}
\begin{barticle}
\bauthor{\bsnm{Hubert}, \binits{L.}},
\bauthor{\bsnm{Arabie}, \binits{P.}}:
\batitle{Comparing partitions}.
\bjtitle{Journal of Classification}
\bvolume{2},
\bfpage{193}--\blpage{218}
(\byear{1985})
\end{barticle}
\endbibitem

\bibitem[\protect\citeauthoryear{Kaufman and
  Rousseeuw}{2009}]{kaufman1990rousseeuw}
\begin{bbook}
\bauthor{\bsnm{Kaufman}, \binits{L.}},
\bauthor{\bsnm{Rousseeuw}, \binits{P.J.}}:
\bbtitle{Finding Groups in Data: an Introduction to Cluster Analysis}.
\bpublisher{John Wiley \& Sons},
\blocation{New York}
(\byear{2009})
\end{bbook}
\endbibitem

\bibitem[\protect\citeauthoryear{Romary et~al.}{2015}]{romary:2015}
\begin{barticle}
\bauthor{\bsnm{Romary}, \binits{T.}},
\bauthor{\bsnm{Ors}, \binits{F.}},
\bauthor{\bsnm{Rivoirard}, \binits{J.}},
\bauthor{\bsnm{Deraisme}, \binits{J.}}:
\batitle{Unsupervised classification of multivariate geostatistical data: Two
  algorithms}.
\bjtitle{Computers \& Geosciences}
\bvolume{85},
\bfpage{96}--\blpage{103}
(\byear{2015})
\end{barticle}
\endbibitem

\bibitem[\protect\citeauthoryear{Wang and Xu}{2017}]{wang2017fast}
\begin{barticle}
\bauthor{\bsnm{Wang}, \binits{X.-F.}},
\bauthor{\bsnm{Xu}, \binits{Y.}}:
\batitle{Fast clustering using adaptive density peak detection}.
\bjtitle{Statistical {M}ethods in {M}edical {R}esearch}
\bvolume{26},
\bfpage{2800}--\blpage{2811}
(\byear{2017})
\end{barticle}
\endbibitem

\bibitem[\protect\citeauthoryear{Bouveyron et~al.}{2015}]{Bouveyron:2015}
\begin{barticle}
\bauthor{\bsnm{Bouveyron}, \binits{C.}},
\bauthor{\bsnm{C{\^o}me}, \binits{E.}},
\bauthor{\bsnm{Jacques}, \binits{J.}}:
\batitle{The discriminative functional mixture model for a comparative analysis
  of bike sharing systems}.
\bjtitle{The {A}nnals of {A}pplied {S}tatistics}
\bvolume{9},
\bfpage{1726}--\blpage{1760}
(\byear{2015})
\end{barticle}
\endbibitem

\bibitem[\protect\citeauthoryear{Bouveyron and Jacques}{2011}]{Bouveyron:2011}
\begin{barticle}
\bauthor{\bsnm{Bouveyron}, \binits{C.}},
\bauthor{\bsnm{Jacques}, \binits{J.}}:
\batitle{Model-based clustering of time series in group-specific functional
  subspaces}.
\bjtitle{{A}dvances in {D}ata {A}nalysis and {C}lassification}
\bvolume{5},
\bfpage{281}--\blpage{300}
(\byear{2011})
\end{barticle}
\endbibitem

\bibitem[\protect\citeauthoryear{Oliver and Webster}{1989}]{Oliver:1989}
\begin{barticle}
\bauthor{\bsnm{Oliver}, \binits{M.}},
\bauthor{\bsnm{Webster}, \binits{R.}}:
\batitle{A geostatistical basis for spatial weighting in multivariate
  classification}.
\bjtitle{Mathematical Geology}
\bvolume{21},
\bfpage{15}--\blpage{35}
(\byear{1989})
\end{barticle}
\endbibitem

\bibitem[\protect\citeauthoryear{Eilers et~al.}{2008}]{eilers2008modulation}
\begin{barticle}
\bauthor{\bsnm{Eilers}, \binits{P.H.}},
\bauthor{\bsnm{Gampe}, \binits{J.}},
\bauthor{\bsnm{Marx}, \binits{B.D.}},
\bauthor{\bsnm{Rau}, \binits{R.}}:
\batitle{Modulation models for seasonal time series and incidence tables}.
\bjtitle{Statistics in Medicine}
\bvolume{27},
\bfpage{3430}--\blpage{3441}
(\byear{2008})
\end{barticle}
\endbibitem

\bibitem[\protect\citeauthoryear{Watanabe and Opper}{2010}]{watanabe:2010}
\begin{barticle}
\bauthor{\bsnm{Watanabe}, \binits{S.}},
\bauthor{\bsnm{Opper}, \binits{M.}}:
\batitle{Asymptotic equivalence of bayes cross validation and widely applicable
  information criterion in singular learning theory.}
\bjtitle{Journal of Machine Learning Research}
\bvolume{11},
\bfpage{3571}--\blpage{3594}
(\byear{2010})
\end{barticle}
\endbibitem

\bibitem[\protect\citeauthoryear{{de la Hoz} et~al.}{2018}]{delaHoz:2018}
\begin{barticle}
\bauthor{\bsnm{{de la Hoz}}, \binits{C.F.}},
\bauthor{\bsnm{Ramos}, \binits{E.}},
\bauthor{\bsnm{Puente}, \binits{A.}},
\bauthor{\bsnm{M{\'e}ndez}, \binits{F.}},
\bauthor{\bsnm{Men{\'e}ndez}, \binits{M.}},
\bauthor{\bsnm{Juanes}, \binits{J.A.}},
\bauthor{\bsnm{Losada}, \binits{{\'I}.J.}}:
\batitle{Ecological typologies of large areas. an application in the
  {M}editerranean {S}ea}.
\bjtitle{Journal of Environmental Management}
\bvolume{205},
\bfpage{59}--\blpage{72}
(\byear{2018})
\end{barticle}
\endbibitem

\bibitem[\protect\citeauthoryear{D'Ortenzio and
  Ribera~d'Alcal{\`a}}{2008}]{d2008trophic}
\begin{barticle}
\bauthor{\bsnm{D'Ortenzio}, \binits{F.}},
\bauthor{\bsnm{Ribera~d'Alcal{\`a}}, \binits{M.}}:
\batitle{On the trophic regimes of the {M}editerranean {S}ea: a satellite
  analysis.}
\bjtitle{Biogeosciences Discussions}
\bvolume{5},
\bfpage{139}--\blpage{148}
(\byear{2008})
\end{barticle}
\endbibitem

\bibitem[\protect\citeauthoryear{Bouzinac et~al.}{2003}]{bouzinac:2003}
\begin{barticle}
\bauthor{\bsnm{Bouzinac}, \binits{C.}},
\bauthor{\bsnm{Font}, \binits{J.}},
\bauthor{\bsnm{Johannessen}, \binits{J.}}:
\batitle{Annual cycles of sea level and sea surface temperature in the western
  {Mediterranean} {S}ea}.
\bjtitle{Journal of Geophysical Research: Oceans}
\bvolume{108}(\bissue{C3}),
\bfpage{3059}
(\byear{2003})
\end{barticle}
\endbibitem

\bibitem[\protect\citeauthoryear{Cutroneo and Capello}{2023}]{cutroneo2023cold}
\begin{barticle}
\bauthor{\bsnm{Cutroneo}, \binits{L.}},
\bauthor{\bsnm{Capello}, \binits{M.}}:
\batitle{The cold waters in the port of {G}enoa ({NW} {M}editerranean {S}ea)
  during the marine heatwave in summer 2022}.
\bjtitle{Journal of Marine Science and Engineering}
\bvolume{11},
\bfpage{1568}
(\byear{2023})
\end{barticle}
\endbibitem

\bibitem[\protect\citeauthoryear{Shaltout and Omstedt}{2014}]{shaltout:2014}
\begin{barticle}
\bauthor{\bsnm{Shaltout}, \binits{M.}},
\bauthor{\bsnm{Omstedt}, \binits{A.}}:
\batitle{Recent sea surface temperature trends and future scenarios for the
  {Mediterranean} {Sea}}.
\bjtitle{Oceanologia}
\bvolume{56},
\bfpage{411}--\blpage{443}
(\byear{2014})
\end{barticle}
\endbibitem

\bibitem[\protect\citeauthoryear{Cade and Noon}{2003}]{cade:2003}
\begin{barticle}
\bauthor{\bsnm{Cade}, \binits{B.S.}},
\bauthor{\bsnm{Noon}, \binits{B.R.}}:
\batitle{A gentle introduction to quantile regression for ecologists}.
\bjtitle{Frontiers in Ecology and the Environment}
\bvolume{1},
\bfpage{412}--\blpage{420}
(\byear{2003})
\end{barticle}
\endbibitem

\bibitem[\protect\citeauthoryear{Kim and Oh}{2020}]{kim-oh:2020}
\begin{barticle}
\bauthor{\bsnm{Kim}, \binits{J.}},
\bauthor{\bsnm{Oh}, \binits{H.-S.}}:
\batitle{Pseudo-quantile functional data clustering}.
\bjtitle{Journal of Multivariate Analysis}
\bvolume{178},
\bfpage{104626}
(\byear{2020})
\end{barticle}
\endbibitem

\bibitem[\protect\citeauthoryear{Bondell
  et~al.}{2010}]{bondell-reich-huixia:2010}
\begin{barticle}
\bauthor{\bsnm{Bondell}, \binits{H.D.}},
\bauthor{\bsnm{Reich}, \binits{B.J.}},
\bauthor{\bsnm{Wang}, \binits{H.}}:
\batitle{Noncrossing quantile regression curve estimation}.
\bjtitle{Biometrika}
\bvolume{97},
\bfpage{825}--\blpage{838}
(\byear{2010})
\end{barticle}
\endbibitem

\end{thebibliography}

\newpage
\begin{appendices}

\begin{center}
\textbf{\Large{Supplementary material for\\ ``Spatial quantile clustering of climate data''}}	\label{secA1}    
\end{center}

\vskip 1cm
\setcounter{table}{0}
\setcounter{figure}{0}

\setcounter{equation}{0}
\setcounter{section}{0}
\section{Conditional distributions in the MCMC algorithm}\label{sec:cond_dist}
We  consider the full set of observations $\boldsymbol{y}=\{\boldsymbol{y}_{1}^\prime,\ldots,\boldsymbol{y}_{N}^\prime\} $  of $ N $   vectors $\boldsymbol{y}_i=(y_{i1},...,y_{i \,T})'$ of independent observations. Each observation  comes from the model
\begin{eqnarray*}
	y_{it}=\boldsymbol{\psi}(t)^\prime\boldsymbol{\gamma}_{c_i}+\tau\sigma_{c_i} w_{it}+\omega\sigma_{c_i} \sqrt{w_{it}}\,\nu_{it}
\end{eqnarray*}
where 
$\tau=(1-2p)/\{p(1-p)\}$, $\omega^2={2}/\{p(1-p)\}$ where  $ \nu_{it}\sim\mathcal{N}(0,1)$ and $ w_{it}\sim \mathcal{E}(1) $ are mutually independent random variables.

The pmf of the random vector 
$\boldsymbol{c}=(\boldsymbol{c}_{1},\ldots,\boldsymbol{c}_{N})^\prime $
(a Potts model) is given by
\paolo{
\begin{eqnarray*}
	f(\boldsymbol{c}|\boldsymbol{\alpha},\beta)
	&=& \frac{\exp\left\{\sum_{k=1}^K  \left(\alpha_k n_k+
	\beta_k\sum_{i=1}^N n_{k,i}/2\right)\right\}}{Z(\boldsymbol{\alpha},\boldsymbol{\beta})},
\end{eqnarray*}
where $n_{k,i}=\sum_{j \in \partial_i} I(c_i=k)I(c_j=k)$ is the number of the neighbors of the site $i$ with  membership equal to $k$.}

The joint density of $ \boldsymbol{y}_i$ given $ \boldsymbol{w}_i=(w_{i1},...,w_{iT})' $ and $c_i$ is
\begin{eqnarray*}
	f(\boldsymbol{y}_i|c_i,\boldsymbol{\gamma}_{c_i},\sigma^2_{c_i},\boldsymbol{w}_i)\propto {(\sigma_{c_i}^2)}^{-{T}/{2}}\prod_{t=1}^{T} w_{it}^{-{1}/{2}}\exp\left\{-\frac{1}{2}\sum_{t=1}^{T}\frac{(y_{it}-\boldsymbol{\psi}(t)^\prime\boldsymbol{\gamma}_{c_i}-\tau\sigma_{c_i} w_{it})^{2}}{\sigma^2_{c_i}\omega^{2}w_{it}}\right\}
\end{eqnarray*}
with $\boldsymbol{\psi}(t)=(\psi_0(t),\ldots,\psi_J(t)^\prime$ and $\boldsymbol{\gamma}_{c_i}=(\gamma_{c_i,0},\ldots,\gamma_{c_i,J})^\prime$. 

We want to make inference on the model parameters $\boldsymbol{\theta}=(\boldsymbol{\alpha},\boldsymbol{\beta},\boldsymbol{\gamma}^\prime,\boldsymbol{\sigma^2}^\prime)'$, with $\boldsymbol{\gamma}=(\boldsymbol{\gamma}_1^\prime,\ldots,\boldsymbol{\gamma}_K^\prime)^\prime$ and $\boldsymbol{\sigma^2}=(\sigma^2_1,\ldots,\sigma^2_K)$.

The prior distribution for $\boldsymbol{\theta}$
 is given by $
 	\pi(\boldsymbol{\theta})=\pi(\boldsymbol{\alpha})\times \pi({\boldsymbol{\beta}})\times \pi(\boldsymbol{\gamma})\times
 	\pi(\boldsymbol{\sigma}^2)
 $
where
\noindent $	 \pi(\boldsymbol{\alpha})=\prod_{k=2}^K	\pi({\alpha}_{k})\propto \prod_{k=2}^K
\exp\left\{-\frac{\alpha_k^2}{2a}
\right\}
$, $a>0$;  

\noindent$\pi(\boldsymbol{\beta})=\paolo{\prod_{k=1}^K I_{(0,b)}(\beta_k)}$, $b>0$

\noindent $	 \pi(\boldsymbol{\gamma})=\prod_{k=1}^K	\pi(\boldsymbol{\gamma}_{k})\propto \prod_{k=1}^K
\exp\left\{-\frac{1}{2g}
\boldsymbol{\gamma}_{k}^\prime\boldsymbol{\gamma}_{k}\right\}
$, $g>0$;

\noindent $\pi(\boldsymbol{\sigma}^2)=\prod_{k=1}^K	\pi(\sigma_{k}^2)\propto \prod_{k=1}^K {(\sigma^2_{k})}^{-s_{0}-1}\exp\left(-{d_{0}}/{\sigma^2_{k}}\right)
$, $s_0>0$ and $d_0>0$ 


Applying the Bayes theorem we obtain the  posterior distribution as

 \begin{eqnarray*}
	\pi(\boldsymbol{\alpha},\boldsymbol{\beta},\boldsymbol{c},\boldsymbol{\gamma},\boldsymbol{\sigma^2}, \boldsymbol{w}|\boldsymbol{y})&\propto&\prod_{i=1}^{N} 
	f(\boldsymbol{y}_i|c_i,\boldsymbol{\gamma}_{c_i},\sigma^2_{c_i}, \boldsymbol{w}_i)f(\boldsymbol{w}_{i})
	\times f(\boldsymbol{c}|\boldsymbol{\alpha},\boldsymbol{\beta})\times \nonumber\\
	&&\prod_{k=1}^K\pi(\boldsymbol{\gamma}_{k})\times \pi(\sigma^2_{k}) \times\pi(\boldsymbol{\alpha})\times \pi(\boldsymbol{\beta}) 
\end{eqnarray*}
 
\subsubsection*{Conditional distribution  for $\boldsymbol{\gamma}$}
\noindent

We have
$
\pi(\boldsymbol{\gamma}|\boldsymbol{\alpha},\boldsymbol{\beta},\boldsymbol{c},\boldsymbol{\sigma^2},\boldsymbol{w},\boldsymbol{y})=\prod_{k=1}^K \pi(\boldsymbol{\gamma}_k|{\sigma}^2_k,\boldsymbol{c},\boldsymbol{w},\boldsymbol{y})$,
where the full conditional posterior distribution of the quantile regression parameters $ \boldsymbol{\gamma}_{c_{i}} $ is obtained as follows
\begin{eqnarray*}
\pi(\boldsymbol{\gamma}_k|{\sigma}^2_k,\boldsymbol{c},\boldsymbol{w},\boldsymbol{y})
	&\propto& \exp\left\{-\frac{1}{2g}\boldsymbol{\gamma}_{k}^\prime\boldsymbol{\gamma}_{k}\right\}\times\\
&&	\prod_{i=1}^{N}
	\left[ \exp\left\{-\frac{1}{2\sigma^2_{c_{i}}\omega^{2}}\sum_{t=1}^{T}\frac{\left(u_{it}-\boldsymbol{\psi}(t)^\prime\boldsymbol{\gamma}_{c_{i}}\right)^{2}}{w_{it}}\right\}	\right]^{I(c_i=k)}
	\nonumber\\
\end{eqnarray*}	
with $u_{it}=y_{it}-\tau\sigma_{c_i}w_{it}$.

The previous formula can be written in compact form using the matrix notation, namely
\begin{eqnarray*}
	\pi(\boldsymbol{\gamma}_k|\sigma^2_k,\boldsymbol{c},\boldsymbol{w},\boldsymbol{y})
	&\propto&\exp\left\{-\frac{1}{2g}\boldsymbol{\gamma}_{k}^\prime\boldsymbol{\gamma}_{k}\right\}\times\exp\left\{-\frac{1}{2}\sum_{i=1}^{N}(\boldsymbol{u}_{i}-\boldsymbol{\Psi}\boldsymbol{\gamma}_{c_{i}})^\prime\boldsymbol{W}_{i}(\boldsymbol{u}_{i}-\boldsymbol{\Psi}\boldsymbol{\gamma}_{c_{i}})I(c_i=k)\right\}\\
	&\propto& 	\exp\left\{-\frac{1}{2}\left[\boldsymbol{\gamma}_{k}'\left(\frac{1}{g}\boldsymbol{I}+\sum_{i=1}^{N}\boldsymbol{\Psi}^\prime\boldsymbol{W}_{i}\boldsymbol{\Psi} I(c_i=k)\right)\boldsymbol{\gamma}_{k}-\right.\right.\\
	&&\left.\left.
 2\boldsymbol{\gamma}_{k}'\left(\sum_{i=1}^{N}\boldsymbol{\Psi}'\boldsymbol{W}_{i}\boldsymbol{u}_{i}I(c_i=k)\right)\right]\right\}
\end{eqnarray*}	
Here $\boldsymbol{u}_i=(u_{i1},\ldots,u_{iT})^\prime$, $\boldsymbol{W}_{i} $ is a diagonal matrix with entries $ [\sigma^2_{c_{i}}\omega^{2} w_{it}]^{-1} $, $ \boldsymbol{\Psi}=(\boldsymbol{\psi}(1),\ldots,\boldsymbol{\psi}(T))^\prime $ is the matrix of covariates.
Thus,  the conditional posterior density of $ \boldsymbol{\gamma}_{k} $ is
proportional to  the  density of a multivariate Gaussian distribution  with vector mean

$$
\boldsymbol{m}_{k}=\left(\frac{1}{g}\boldsymbol{I}+\sum_{i=1}^{N}\boldsymbol{\Psi}^\prime\boldsymbol{W}_{i} \boldsymbol{\Psi} I(c_i=k)\right)^{-1}\left(\sum_{i=1}^{N}\boldsymbol{\Psi}^\prime\boldsymbol{W}_{i} \mathbf{u}_{i}I(c_i=k)\right)
$$
and covariance matrix
$$
\boldsymbol{S}_k=\left(\frac{1}{g}\boldsymbol{I}+\sum_{i=1}^{N}\boldsymbol{\Psi}^\prime\boldsymbol{W}_{i} \boldsymbol{\Psi} I(c_i=k)\right)^{-1}
$$

\subsubsection*{Full conditional  for $\boldsymbol{\sigma}^2$}
\noindent
We note  that 
\begin{eqnarray*}
	\pi(\boldsymbol{\sigma}^2|\boldsymbol{\alpha},\boldsymbol{\beta},\boldsymbol{\gamma},\boldsymbol{c},\boldsymbol{w},\boldsymbol{y}) 
		\propto\prod_{k=1}^K \pi(\sigma^2_k|\boldsymbol{\gamma}_k,\boldsymbol{c},\boldsymbol{w},\boldsymbol{y})
\end{eqnarray*}
To obtain the full conditional posterior of $ \sigma^2_{k} $ we proceed as follows
\begin{eqnarray}
\pi(\sigma_{k}^2|\boldsymbol{\gamma}_{k},\boldsymbol{c},\boldsymbol{w},\boldsymbol{y})
&\propto& (\sigma^2_{k})^{-s_{0}-1}\exp\left(-\frac{d_{0}}{\sigma^2_{k}}\right)\times
\nonumber\\
&&\prod_{i=1}^{N}
\left[ {(\sigma^2_{c_{i}})}^{-{T}/{2}}\exp\left\{-\frac{1}{2\sigma^2_{c_{i}}\omega^{2}}\sum_{t=1}^{T}\frac{\left(u_{it}-\boldsymbol{\psi}(t)^\prime \boldsymbol{\gamma}_{c_{i}}\right)^{2}}{w_{it}}\right\}
\right]^{I(c_i=k)}\nonumber\\
&\propto& {(\sigma_{k}^2)}^{-(s_{0}+n_{k}T/2)-1}\times \nonumber\\ 
&&\exp\left[
-\frac{1}{\sigma^2_{k}}\left[d_{0}+\frac{1}{2\omega^{2}}\sum_{i=1}^{N}\left\{\sum_{t=1}^{T}\frac{\left(u_{it}-\boldsymbol{\psi}(t)^\prime \boldsymbol{\gamma}_{c_{i}}\right)^{2}}{w_{it}}\right\}I(c_i=k)\right]\right]\nonumber\\ &&\,\label{eq:s7}
\end{eqnarray}
where $n_{k}=\sum_{i=1}^{N}I(c_i=k)$, which  represents the number of vectors of $ \boldsymbol{y} $ with membership $ k $.

Therefore, \eqref{eq:s7} is the distribution of an inverse Gamma random variable with  shape  parameter and scale parameter, $$a=s_{0}+n_{k}T/2,\qquad 
 b=d_{0}+\frac{1}{2\omega^{2}}\sum_{i=1}^{N}\left\{\sum_{t=1}^{T}\frac{\left(u_{it}-\boldsymbol{\psi}(t)^\prime \boldsymbol{\gamma}_{c_{i}}\right)^{2}}{w_{it}}\right\}I(c_i=k).$$
  We note that random samples from inverse Gamma distribution can be drawn from a  Gamma distribution observing the relationship that  if $G \sim \mathrm{Gamma}(a, 1/b)$  then $1/G \sim \mathrm{InvGamma}(a, b)$.

\subsubsection*{Conditional distribution  for $\boldsymbol{w}$}
\noindent
We note  that 
\begin{eqnarray*}
\pi(\boldsymbol{w}|\boldsymbol{\alpha},\boldsymbol{\beta},\boldsymbol{\gamma},\boldsymbol{\sigma}^2,\boldsymbol{c},\boldsymbol{y})&\propto& \prod_{i=1}^{N}
f(\boldsymbol{w}_{i}|\boldsymbol{\gamma}_{c_{i}},\sigma^2_{c_{i}},c_{i},\boldsymbol{y}_{i})=\prod_{i=1}^{N}\prod_{t=1}^T
f({w}_{it}|\boldsymbol{\gamma}_{c_{i}},\sigma^2_{c_{i}},c_{i},y_{it})
\end{eqnarray*}
and
\begin{eqnarray*}
	f({w}_{it}|\boldsymbol{\gamma}_{c_{i}},\sigma^2_{c_{i}},c_{i},y_{it})&\propto& f(y_{it}|\boldsymbol{\gamma}_{c_{i}},\sigma^2_{c_{i}},c_{i},w_{it})\times f(w_{it})\\
	&\propto& w_{it}^{-{1}/{2}}\exp\left\{-\frac{1}{2\sigma^2_{c_{i}}\omega^{2}w_{it}}\left(y_{it}-\boldsymbol{\psi}(t)^\prime \boldsymbol{\gamma}_{c_{i}}- \tau \sigma_{c_i} w_{it}\right)^{2}-w_{it}\right\}\\
	&\propto& w_{it}^{-{1}/{2}}\exp\left[-\frac{1}{2\sigma^2_{c_{i}}\omega^{2}w_{it}}\left\{(y_{it}-\boldsymbol{\psi}(t)^\prime \boldsymbol{\gamma}_{c_{i}})^{2}+\tau^{2}\sigma^2_{c_{i}}w_{it}^{2}\right\}-w_{it}\right]\\
	&\propto& w_{it}^{-{1}/{2}}\exp\left\{-\frac{1}{2\sigma^2_{c_{i}}\omega^{2}w_{it}}(y_{it}-\boldsymbol{\psi}(t)^\prime \boldsymbol{\gamma}_{c_{i}})^{2}-\frac{\tau^{2}w_{it}}{2\omega^{2}}-w_{it}\right\}\\
	&\propto& w_{it}^{-{1}/{2}}\exp\left\{-\frac{1}{2\sigma^2_{c_{i}}\omega^{2}w_{it}}(y_{it}-\boldsymbol{\psi}(t)^\prime \boldsymbol{\gamma}_{c_{i}})^{2}-\frac{w_{it}}{2\omega^{2}}(\tau^{2}+2\omega^{2})\right\}\\
		&\propto& w_{it}^{-{1}/{2}}\exp\left[-\frac{1}{2}\left\{\frac{\tau^{2}+2\omega^{2}}{\omega^{2}}w_{it}+\frac{(y_{it}-\boldsymbol{\psi}(t)^\prime \boldsymbol{\gamma}_{c_{i}})^{2}}{\sigma^2_{c_{i}}\omega^{2}w_{it}}
\right\}\right]\\
\end{eqnarray*}
This expression resembles a Generalized Inverse Gaussian distribution $ \operatorname{GIG}(d,a,b) $ with the density
\begin{eqnarray*}
	f(w;a,b,d)&\propto& w^{(d-1)}\exp\left\{-\frac{1}{2}\left(aw+\frac{b}{w}\right)\right\}
\end{eqnarray*}
where $d=0.5$, $\displaystyle	a=\frac{\tau^{2}+2\omega^{2}}{\omega^{2}}$, $\displaystyle b=\frac{(y_{it}-\boldsymbol{\psi}(t)^\prime \boldsymbol{\gamma}_{c_{i}})^{2}}{\sigma^2_{c_{i}}\omega^{2}}$.
\newpage
\section{Selection of the number of elements for the B-spline basis}
\begin{table}[h]
\caption{Overall $\operatorname{BIC}$ for different values of $J_1$, $J_2$, and $J_3$.}
\begin{tabular}{lccccccc}
\hline
 & \multicolumn{5}{c}{$J_3$}\\
$J_2=4$  & 4 & 5 & 6 & 7 & 8\\
  \hline
$J_1=4$ & 5129 & 7626 & 9791 & 12092 & 14527 \\ 
$J_1=5$ &5796 & 8223 & 10419 & 12664 & 15095 \\ 
$J_1=6$ &4802 & 7179 & 9348 & 11450 & 13638 \\ 
$J_1=7$ &5074 & 7494 & 9632 & 11721 & 13852 \\ 
$J_1=8$ &5614 & 8042 & 10190 & 12315 & 14459 \\ 
  \hline
 & \multicolumn{5}{c}{$J_3$}\\
$J_2=5$  & 4 & 5 & 6 & 7 & 8\\
  \hline
$J_1=4$ &-9044 & -6861 & -5066 & -2993 & -737 \\ 
$J_1=5$ &-8385 & -6204 & -4435 & -2416 & -120 \\ 
$J_1=6$ &-9998 & -7865 & -6025 & -4044 & -1836 \\ 
$J_1=7$ &-9996 & -7847 & -5990 & -4094 & -2105 \\ 
$J_1=8$ &-9555 & -7426 & -5575 & -3723 & -1745 \\ 
  \hline
 & \multicolumn{5}{c}{$J_3$}\\
$J_2=6$  & 4 & 5 & 6 & 7 & 8\\
  \hline
$J_1=4$ &-15650 & -13529 & -11702 & -9573 & -7432 \\ 
$J_1=5$ &-14936 & -12847 & -11019 & -8943 & -6807 \\ 
$J_1=6$ &-16313 & -14273 & -12358 & -10281 & -8277 \\ 
$J_1=7$ &-16492 & -14460 & -12573 & -10668 & -8736 \\ 
$J_1=8$ &-16077 & -14065 & -12162 & -10251 & -8358 \\
  \hline
 & \multicolumn{5}{c}{$J_3$}\\
$J_2=7$  & 4 & 5 & 6 & 7 & 8\\
  \hline
$J_1=4$ &-21116 & -19031 & -17445 & -15082 & -12877 \\ 
$J_1=5$ &-20299 & -18240 & -16678 & -14304 & -12083 \\ 
$J_1=6$ &-21213 & -19147 & -17541 & -15076 & -12865 \\ 
$J_1=7$ &-21716 & -19716 & -18138 & -15778 & -13607 \\ 
$J_1=8$ &-21459 & -19451 & -17934 & -15579 & -13423 \\
  \hline
 & \multicolumn{5}{c}{$J_3$}\\
$J_2=8$  & 4 & 5 & 6 & 7 & 8\\
  \hline
$J_1=4$ &-23877 & -21808 & -20437 & -18240 & -16081 \\ 
$J_1=5$ &-23084 & -21036 & -19738 & -17507 & -15358 \\ 
$J_1=6$ &-24149 & -22158 & -20856 & -18480 & -16268 \\ 
$J_1=7$ &\textbf{-24547} & -22643 & -21478 & -19185 & -17028 \\ 
$J_1=8$ &-24417 & -22532 & -21357 & -19095 & -16945 \\ 
   \hline
   
\hline
\end{tabular}
\label{tab:BIC}
\end{table}

\newpage
\section{Selection of number of clusters }
\begin{figure}[ht]
  \centering
  \begin{tabular}{cc}
\includegraphics[width=0.45\linewidth]{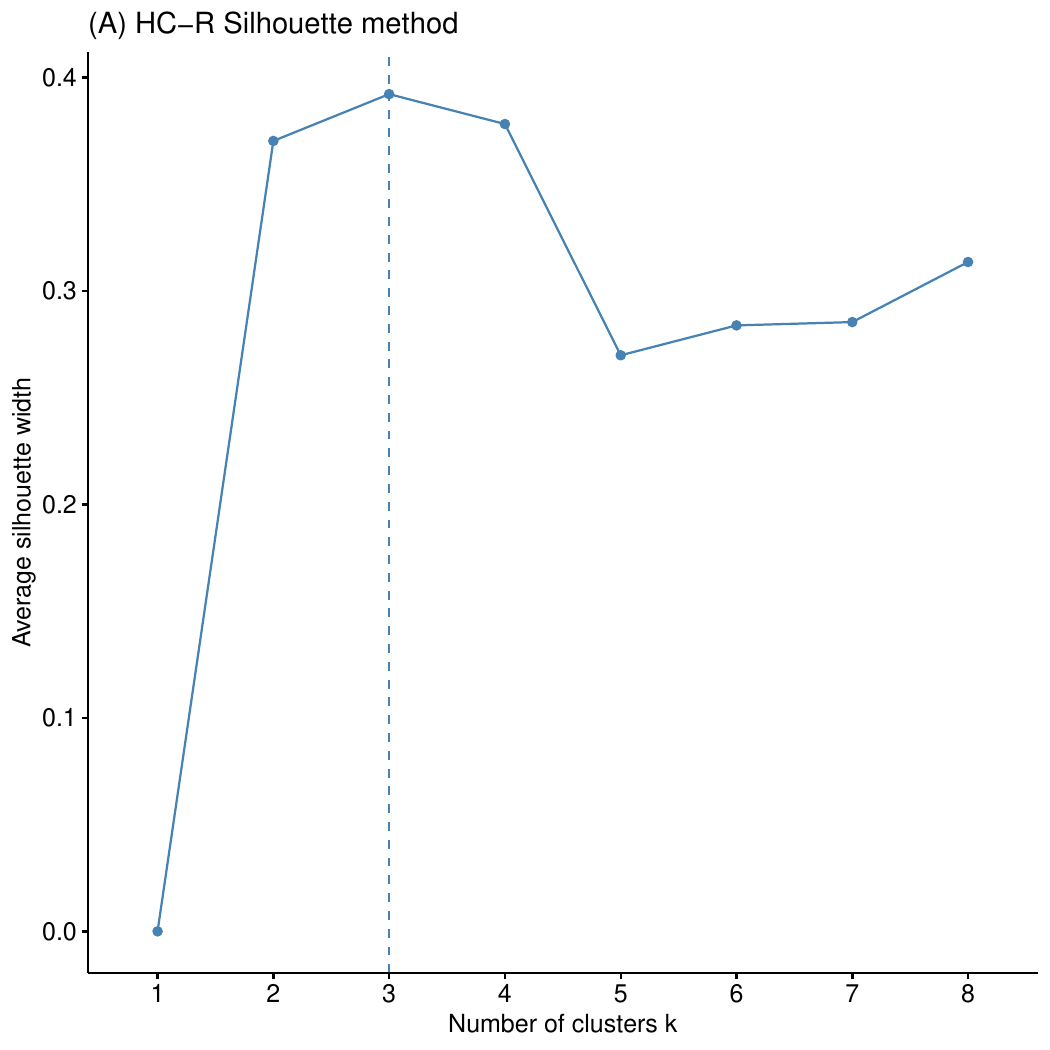}&
\includegraphics[width=0.45\linewidth]{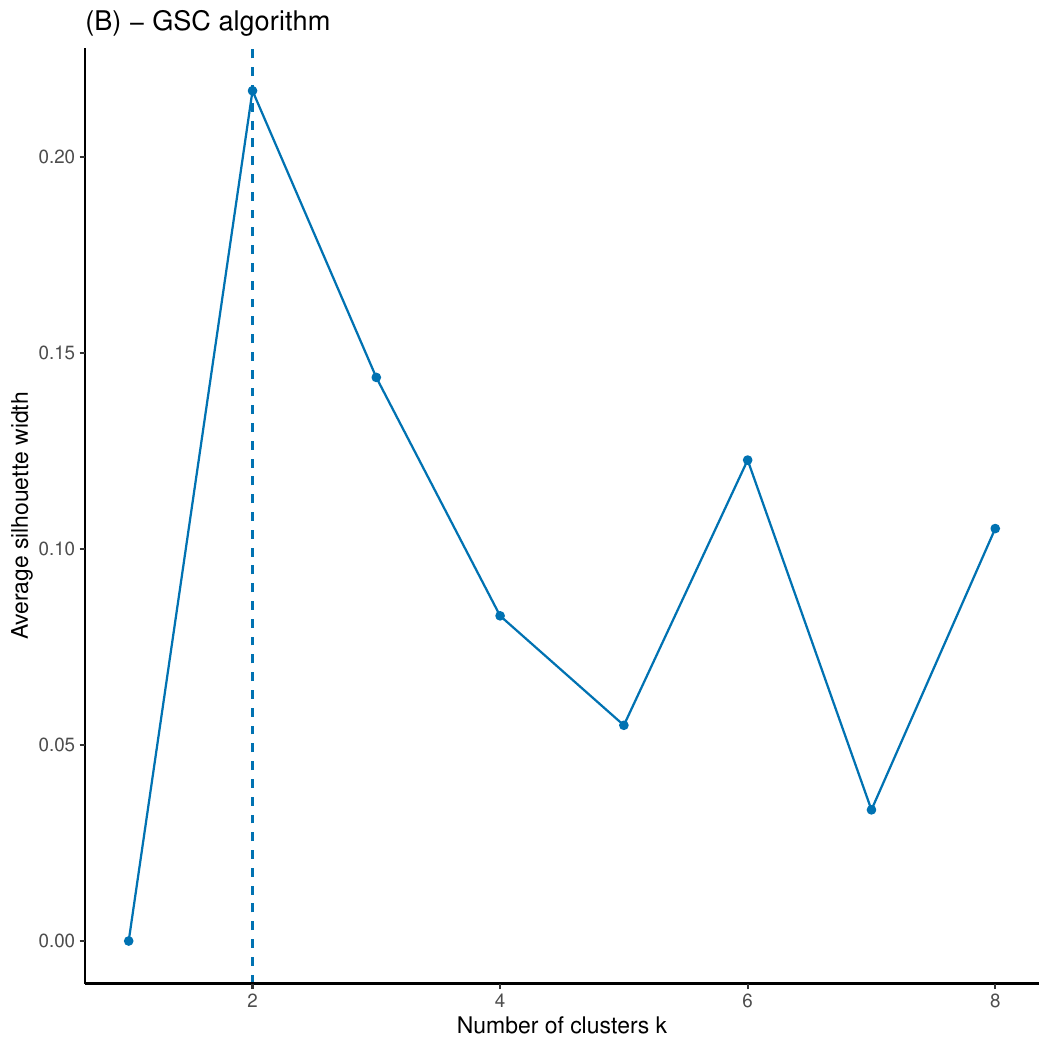}\\
\includegraphics[width=0.45\linewidth]{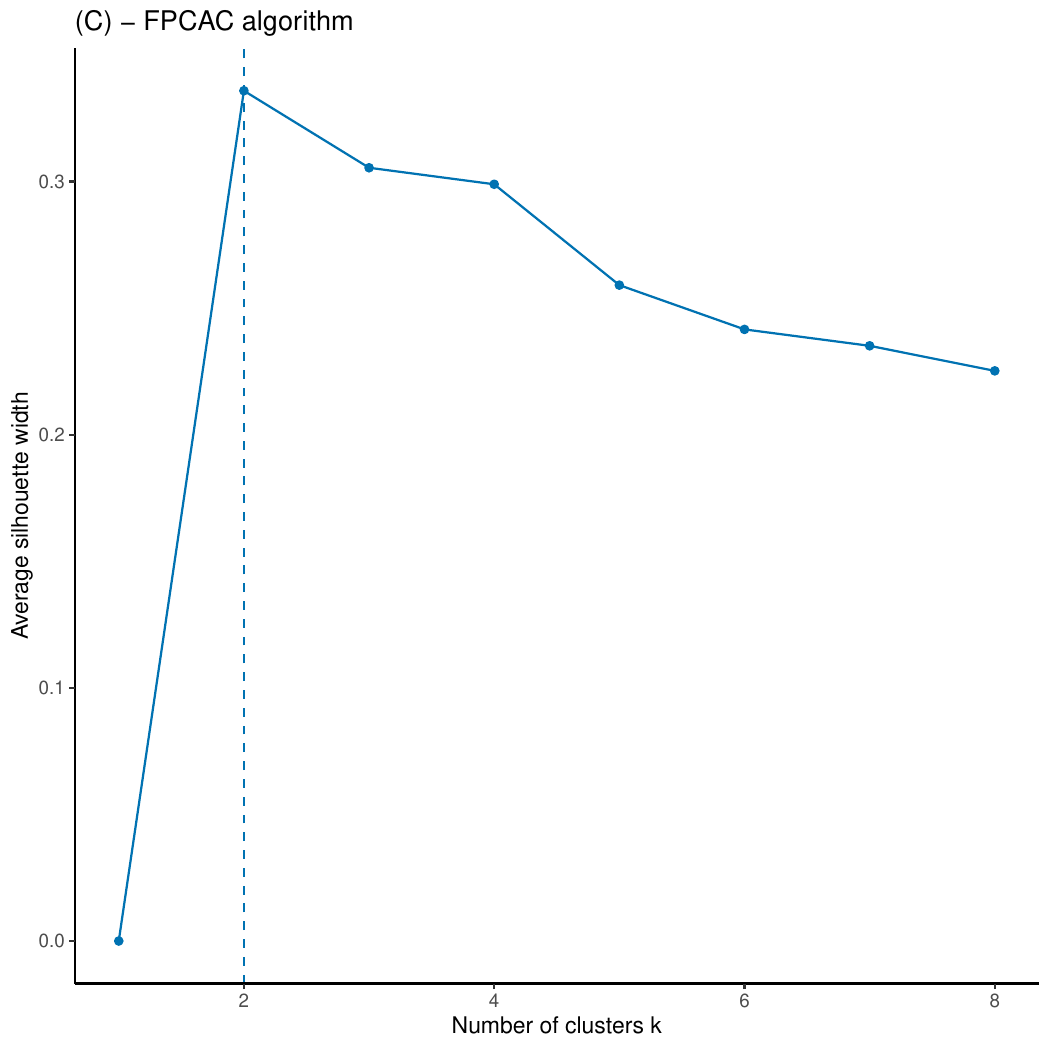}&
\includegraphics[width=0.45\linewidth]{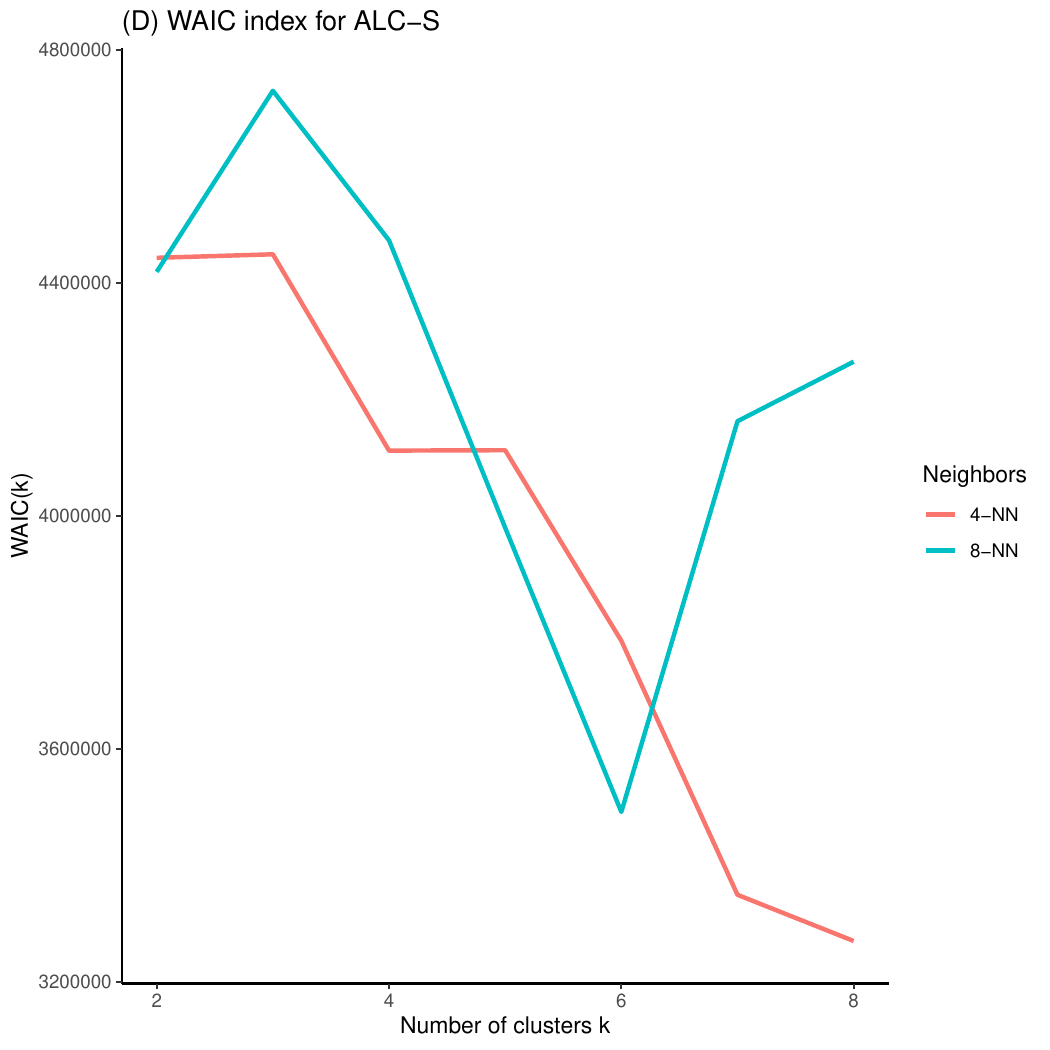}\\
  \end{tabular}
	\caption{Silhouette profile for HC procedures and $\operatorname{WAIC}$ values for ALC-S method with 4 and 8 nearest neighbors \typo{by} number of clusters.\label{figure_waic} }
\end{figure}

\end{appendices}
\end{document}